\documentclass[]{aa}
\usepackage{graphicx}
\usepackage{aalongtable}
\usepackage{txfonts}
\begin{document}
   \title{Brown dwarfs and very low mass stars in the Hyades cluster :
   a dynamically evolved mass function.}

   \author{J. Bouvier\inst{1}, T. Kendall\inst{2}, G. Meeus\inst{3},
     L. Testi\inst{4,5}, E. Moraux\inst{1}, J.R. Stauffer\inst{6},
     D. James\inst{7}, J.-C. Cuillandre\inst{8}, J. Irwin\inst{9},
     M.J. McCaughrean\inst{10}, I. Baraffe\inst{11}, \and
     E. Bertin\inst{12}\fnmsep\thanks{Partly based on observations
     obtained at the Canada-France-Hawaii Telescope (CFHT) which is
     operated by the National Research Council of Canada, the Institut
     National des Science de l'Univers of the Centre National de la
     Recherche Scientifique of France, and the University of
     Hawaii. Partly based on observations made with the Italian
     Telescopio Nazionale Galileo (TNG) operated on the island of La
     Palma by the Fundaci\'on Galileo Galilei of the INAF (Istituto
     Nazionale di Astrofisica) at the Spanish Observatorio del Roque
     de los Muchachos of the Instituto de Astrofisica de Canarias.
     Partly based on observations made with ESO Telescopes at the La Silla
     Observatories. } }

   \offprints{J. Bouvier}

   \institute{Laboratoire d'Astrophysique de Grenoble, Observatoire de
   Grenoble, BP53, F-38041 Grenoble, France
\and
        Centre for Astrophysics Research, University of Hertfordshire,
   College Lane, Hatfield AL10 9AB, UK
\and 
        Astrophysikalisches Institut Potsdam, An der Sternwarte 16, D-14482
   Potsdam, Germany
\and 
        Osservatorio Astrofisico di Arcetri, INAF, Largo E. Fermi 5,
   50125 Firenze, Italy
\and  European Southern Observatory, Karl Schwarzschild Str. 2, 85748
   Garching bei M\"unchen, Germany
\and
        Spitzer Science Center, California Institute of Technology, 1200
   East California Boulevard, Pasadena, CA 91125, USA
\and
        Department of Physics \& Astronomy,  Box 1807 Station B, Vanderbilt
   University, Nashville, TN 37235, USA
\and 
         CFHT Corporation, 65-1238 Mamalahoa Hwy, Kamuela, Hawaii 96743,
   USA  
\and   Harvard-Smithsonian Center for Astrophysics, 60 Garden Street
   MS-16, Cambridge, MA 02138, USA
\and   School of Physics, Stocker Road, Exeter, EX4 4QL, UK 
\and
        CRAL, Ecole Normale Sup\'erieure,  46 all\'ee d'Italie, 69364 Lyon
   Cedex 07, France
\and 
        Institut d'Astrophysique de Paris, 98bis bd Arago, 75014 Paris,
   France
}

   \date{Received; accepted}
   
   \abstract 
       {} 
       {We conducted a search for brown dwarfs (BDs) and very
   low mass (VLM) stars in the 625~Myr-old Hyades cluster in order to
   derive the cluster's mass function across the stellar-substellar
   boundary.} {We performed a deep (I=23, z=22.5) photometric
   survey over 16~deg$^2$ around the cluster center, followed up with
   K-band photometry to measure the proper motion of candidate
   members, and optical and near-IR spectroscopy of probable BD and
   VLM members.} 
       {We report the discovery of the first 2 brown dwarfs
   in the Hyades cluster. The 2 objects have a spectral type early-T
   and their optical and near-IR photometry as well as their proper
   motion are consistent with them being cluster members. According to
   models, their mass is 50 Jupiter masses at an age of 625~Myr. We
   also report the discovery of 3 new very low mass stellar members of
   the cluster, and confirm the membership of 16 others. We combine
   these results with a list of previously known cluster members to
   build the present-day mass function (PDMF) of the Hyades cluster
   from 50 Jupiter masses to 3~M$_\odot$. We find the Hyades PDMF to
   be strongly deficient in very low mass objects and brown dwarfs
   compared to the IMF of younger open clusters such as the
   Pleiades. We interpret this deficiency as the result of dynamical
   evolution over the past few 100~Myr, i.e., the preferential
   evaporation of low mass cluster members due to weak gravitational
   encounters.}
       {We thus estimate that the Hyades cluster currently hosts about
   10-15 brown dwarfs, while its initial substellar population may have
   amounted up to 150-200 members.}

   \keywords{Stars: low-mass, brown dwarfs -- Stars: luminosity function,
     mass function -- (Galaxy:) open clusters and associations: individual:
     Hyades (Melotte 25) } 

\authorrunning{J. Bouvier et al.}

   \maketitle

\section{Introduction}

The determination of the initial mass function (IMF), i.e., the mass
frequency distribution of objects formed in molecular clouds, yields
strong constraints to star formation theories (Padoan \& Nordlund
2002; Bate \& Bonnell 2005; Larson 2005; Jappsen et al. 2005). In the
last 10 years, the IMF has been derived from the most massive stars
down to the substellar domain in a variety of environments : star
forming regions (e.g. Luhman et al. 2003), young open clusters
(e.g. Bouvier et al. 1998) and in the field (e.g. Reid et
al. 1999). As brown dwarfs continuously cool down as they evolve
(Chabrier et al. 2000), they are brighter when younger. Thus, the IMF
could be derived down to a mass of a few Jupiter masses in star
forming regions (Lucas et al. 2006; Caballero et al. 2007) and down to
about 30 Jupiter masses in rich young open clusters (Moraux et
al. 2003, 2007; Barrado et al. 2004; de Wit et al. 2006). While a
large number of field brown dwarfs have also been discovered within a
few tens of parsecs from the Sun, the derivation of the field IMF
below 0.1~M$_\odot$ still remains uncertain due to the lack of
knowledge of the ages of substellar objects (Chabrier 2002; Allen et
al. 2005; Cruz et al. 2007).

In a previous series of paper, we derived the lower mass function of a
number of young open clusters down to about 30~M$_{Jup}$ (see Moraux,
Bouvier \& Clarke 2005 and Bouvier, Moraux \& Stauffer 2005 for short
reviews) and discussed the implications of an apparently universal
cluster mass function for star formation scenarios (Moraux et
al. 2007). Here, we report an extended, deep optical survey of the
Hyades cluster aimed at detecting substellar objects and deriving the
cluster's mass function across the stellar-substellar boundary. The
motivations for this survey are twofold. Firstly, all previous
searches for substellar objects in the Hyades have failed to report
any positive brown dwarf detections (Reid \& Hawley 1999; Gizis, Reid
\& Monet 1999; Dobbie et al. 2002). The lowest mass members reported
so far includes LH~0418+13, with a spectral type M8.5 and an estimated
mass of 0.083~M$_\odot$ (Reid \& Hawley 1999), the
$\simeq$0.081~M$_\odot$ companions of a couple of 0.1~M$_\odot$ Hyades
probable members, LP~415-20 and LP~475-855 (Siegler et al. 2003), and
an unresolved very low mass companion (0.070-0.095~M$_\odot$) in the
spectroscopic binary RHy~403 (Reid \& Mahoney 2000). Yet, the
discovery of brown dwarfs at an age of 625~Myr (Perryman et al. 1998)
would provide a unique benchmark to calibrate the substellar evolution
models. Secondly, the Hyades cluster is dynamically evolved (Adams et
al. 2002). A significant fraction of its initial low mass population
is therefore expected to have drifted away beyond the cluster
boundaries (e.g. Reid 1993). Deriving the lower mass function of such
an evolved cluster would provide a direct measurement of the rate at
which low mass cluster members evaporate and populate the field. In
turn, this measurement would allow us to test the validity of N-body
simulations of the dynamical evolution of young clusters (e.g. Kroupa
1995; Portegies Zwart et al. 2001).

The Hyades cluster (Melotte 25, $\alpha_{2000}$=04$^h$26$^m$54$^s$,
$\delta_{2000}$=+15$\degr$52$\arcmin$; $l$=180.05$\degr$,
$b$=-22.40$\degr$) is the closest rich open cluster to the
Sun. Perryman et al. (1998) derived its main structural and
kinematical properties based on Hipparcos measurements~: a distance of
46.3$\pm$0.27~pc, an age of 625$\pm$50~Myr, a metallicity [Fe/H] of
0.14$\pm$0.05, a present-day total mass of about 400~M$_\odot$, a
tidal radius of 10.3~pc, a core radius of 2.5-3.0~pc and negligible
extinction on the line of sight. The large proper motion of the
cluster ($\mu\simeq$100~mas yr$^{-1}$) can be easily measured from
imaging surveys over a timeframe of only a few years, which helps in
assessing cluster's membership. 

In Section~2, we describe the optical survey we performed over the
central 16~deg$^2$ of the Hyades cluster, as well as follow-up
K-band photometry and both optical and infrared spectroscopy. The
proper motion of optically-selected candidate members is derived from
optical and near-IR images obtained 2 to 3 years apart. In Section~3,
we describe our selection of candidate members combining optical and
inrafred photometry, as well as proper motion measurements and follow
up spectroscopy. We thus report 21 probable members, of which 5 are
new and 2 are the first Hyades brown dwarfs, with an estimated mass of
$\simeq$50~M$_{Jup}$. In Section~4, we discuss the spectral properties
of the newly found low mass Hyades members, and derive a spectral type
of T1 and T2 for the 2 brown dwarf candidates. We proceed in deriving
the present-day mass function of the Hyades cluster from 0.050 to
3.0~M$_\odot$, which we find to be strongly deficient in very low mass
stars and brown dwarfs compared to the mass function of the younger
Pleiades cluster. We discuss this result in the light of N-body
simulations of the dynamical evolution of young open clusters.

\section{Observations and data reduction}

We describe below the deep, wide field optical survey we performed on
the Hyades cluster at CFHT in the I and z bands. We also present
follow up K-band imaging, as well as optical and infrared spectroscopy
obtained for a subset of optically-selected cluster candidate members.

\subsection{The optical survey}

\begin{figure}[t]
   \centering
   \includegraphics[width=9cm]{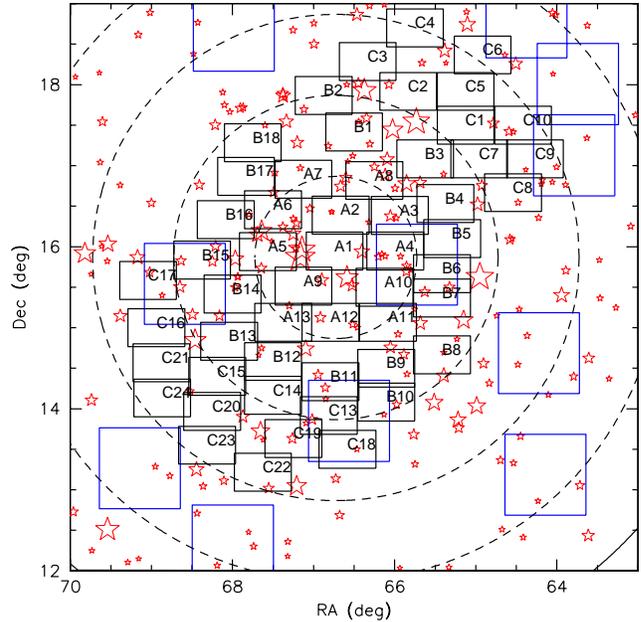}
      \caption{The spatial distribution of CFHT12K Hyades fields is shown
              by rectangles labelled with their names
              (cf. Table~\ref{fieldscoo}). Dashed concentric circles
              are spaced by 1$\degr$ around the cluster center. Star
              symbols (red) are Hyades probable members listed in
              Prosser \& Stauffer's database, with a size related to
              their luminosity. Large (blue) squares indicate the area
              previously surveyed by Dobbie et al. (2002). }
         \label{fields}
   \end{figure}

Wide-field optical images were obtained in the I and z bands with the
CFHT 12K camera, a mosaic of 12 CCD arrays with a pixel size of
0.21$\arcsec$ which provides a FOV of 42$\times$28$\arcmin$
(Cuillandre et al. 2000). The survey was performed during several
service observing runs between September 2002 and January 2003, and
consists of 53 mosaic fields covering a total of 17.3 square degrees
over the cluster area. However, one CCD of the mosaic (CCD05) had
serious cosmetics problems during these runs and yielded useless data,
thus reducing the effective survey area by 1/12. The dates of
observation and the coordinates of the center of the mosaic fields are
listed in Table~\ref{fieldscoo} in Appendix~A. The area covered by the
survey on the sky is illustrated in Figure~\ref{fields}. The survey
extends symmetrically around the cluster's center, along a 4 deg-wide
stripe of constant galactic latitude, and up to 3 degrees away from
the cluster center in galactic longitude.

%----------------------------------------------------------- Fig.1
%Fig.1 = /gagax1/tera/jbouvier/CFHTKP_CD220404/hyades_02B/OB02B/hyades_JRS.ps

%
%______________________________________________________________

Short and long exposures (10s in I and z, 3$\times$300s in I and
3$\times$360s in z) were obtained for each field. This resulted in a
continuously sampled range of I magnitudes from I$\sim$13, the
saturation limit of short exposures, to a detection limit of
I$\sim$24. The survey is at least 90\% complete down to I$\sim$23.0
and z$\sim$22.5, as computed from overlapping 12K fields (see Moraux
et al. 2003), a limit which varies only slightly with seeing
conditions (0.6-0.8$\arcsec$ FWHM, see
Table~\ref{fieldscoo}). Photometric standard fields (Landolt 1992)
were observed several times per night, as part of the continuous
monitoring of the CCD mosaic photometric zero points during service
observing runs.

Individual images were processed through CFHT's Elixir reduction
pipe-line (Magnier \& Cuillandre 2004) including bias subtraction,
flat-fielding, fringe correction, scattered light correction in
z-band, and astrometric calibration (0.8$\arcsec$ rms). Multiple I-
and z-band dithered long exposures were then registered and co-added.
PSF photometry was performed on both short and long exposures with a
modified version of SExtractor (Bertin \& Arnouts 1996) from a PSF
model computed with the PSFEx software (Bertin et al., in prep.). The
resulting catalogues were photometrically calibrated using the I- and
z-band photometric zero points of the corresponding nights derived
from the Landolt standard fields and extinction coefficients $k_I$ and
$k_z$ of 0.04 and 0.03 mag/airmass, respectively. In the z-band, the
photometric zero points are computed from selected Landolt standards
of spectral type A0. A few frames were taken through thin cirrus
absorption (a few 0.01 mag). In this case, additional I and z
exposures were obtained (30s in I, 36s in z) for the same fields on
subsequent photometric nights and used to recalibrate the original
short and long exposure images. 

Note that even though Landolt standards have been used to compute the
photometric zero-points, the resulting I-band magnitudes are not in
the Cousins system. This is due to the different passbands of the
CFHT12K and Cousins filters which induces color terms, especially for
very red objects. These color-terms can be estimated by integrating
both filter passbands over Baraffe et al.'s (1998) model spectra. This
yields : $I_{12K} = I_c$ for $(I_c-K) \leq 2.0$ and $I_{12K} = I_c -
0.034 \times (I_c-K) + 0.068 $ for $2.0 \leq (I_c-K) \leq 6.0$. We have not
applied these color terms on our photometry. Instead, 600~Myr
theoretical isochrones from the Lyon models were specifically computed
for the I- and z-band CFHT filters in order to provide a consistent
comparison between models and observations in color-magnitude
diagrams.

We found slight systematic differences between the photometric zero point
of the various CCDs within the mosaic. The CCD-to-CCD zero point shifts
were empirically estimated from the median (I-z) color of all stellar-like
objects in each CCD, with reference to CCD04. The distribution of color
shifts, computed over all images, has a median value of -0.005 mag and a
rms of 0.05 mag. The CCD-to-CCD zero point shifts were corrected for by
applying them to the z-band magnitude of the objects in the output
photometric catalogues. The I-band magnitude was left unchanged. Small
overlaps between most of the observed fields eventually allowed us to check
the quality of the photometric calibration. This was done by comparing the
pair of magnitude measurements available for every star common to 2
overlapping fields. 
%Figure~\ref{overlaps} shows the difference of these
%pairs of measurements, after residual shifts were corrected for, for all
%stars lying in the overlapping areas of long exposure I and z-band
%exposures of the Hyades A fields (similar results are obtained for Hyades B
%and C fields). 
The rms accuracy of the I-band and z-band photometry is found to be of
order of 0.06 mag for I=18-20 in long exposures, then increases to
0.08 mag in the range I=20-22, and as the signal to noise ratio
decreases further reaches 0.10 mag at I$\sim$22.5.  Similar results,
scaled to the corresponding magnitude range, are obtained for short
exposure images.

%----------------------------------------------------------- Fig.3
%Fig.3 = /home/jbouvier/HyadesABC/paper/CMD

   \begin{figure}[t]
   \centering
   \includegraphics[width=8cm]{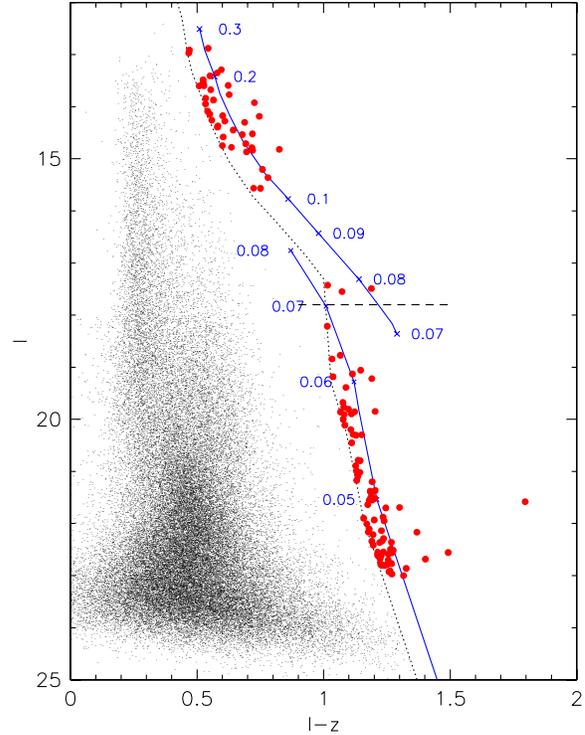}
      \caption{(I, I-z) color-magnitude diagram of stellar like
        objects detected on the Hyades CFHT12K frames. 600 Myr NextGen
        (upper) and Dusty (lower) isochrones, shifted to the Hyades
        distance, are shown as solid lines and labelled with mass
        (M$_\odot$). The horizontal dashed line illustrates the
        expected location of the stellar/substellar boundary along the
        cluster sequence. Hyades candidate members (large red dots)
        were selected to the right of the dotted line running close to
        the isochrones (see text). For clarity, only 1 in 10 objects
        of the background galactic population is shown leftwards of
        the selection line.  }
         \label{CMDiiz}
   \end{figure}
%
%______________________________________________________________

The photometric catalogues originally contained nearly 2.3 million non
saturated objects detected on the short and long exposures. Non
stellar objects (galaxies, bad pixels, cosmics) were rejected from the
photometric catalogues on the basis of the FHWM of their PSF. The
seeing FWHM was measured on each image from plots of the object's
magnitude against PSF FWHM. Stellar-like objects lie along a well
defined vertical line in these plots, at the value of the seeing FWHM
(typically 0.6-0.8$\arcsec$, i.e., 3-4 pixels FWHM). Conservative limits
typically between 50 and 150\% of the seeing FWHM were adopted to
filter out galaxies (larger PSF FWHM) and bad pixels and cosmics
(smaller PSF FWHM). Eventually, the output photometric catalogue
contains over 930,000 non saturated stellar-like objects detected in
both I- and z-bands over the 16 square degrees we surveyed down to a
detection limit of I$\sim$z$\sim$24. The corresponding (I, I-z) color
magnitude diagram (CMD) is shown in Fig~\ref{CMDiiz}. A total of 125
possible Hyades members were selected in this CMD from their location
relative to model isochrones (see Section~3.1).

%----------------------------------------------------------- Fig.4
%Fig.4 = /home/jbouvier/CFHTIR05B/PPM/CROSSCOR

   \begin{figure*}[t]
   \centering
   \includegraphics[width=17cm]{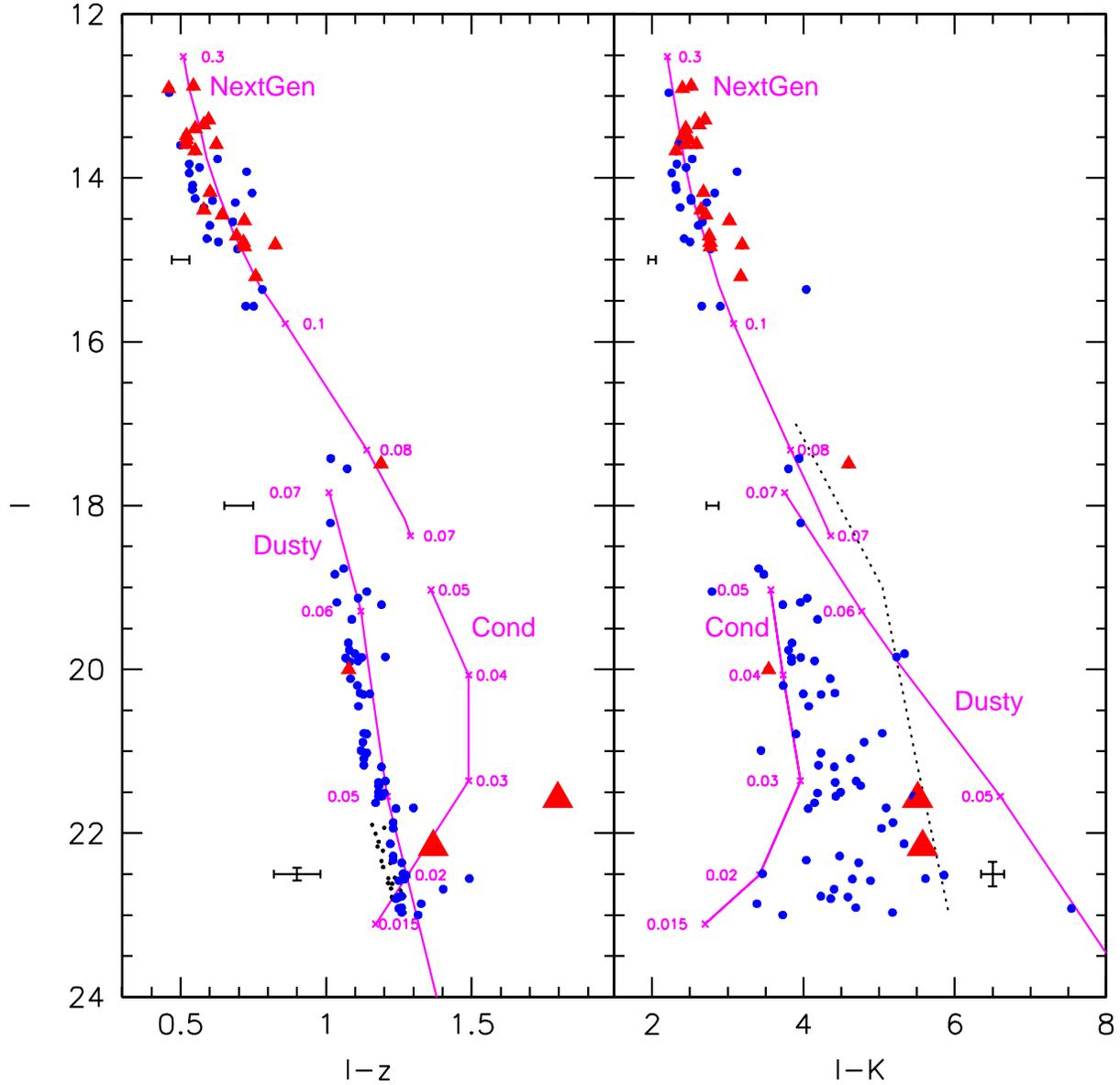}
      \caption{ Left~: (I, I-z) CMD of optically selected
   candidates. Right~: (I, I-K) CMD of optically selected candidated
   followed up with CFHT IR in the K-band. Small dots (left) : 17
   optically selected candidate without follow up IR photometry. Large
   dots : optically selected candidates whose proper motion is
   inconsistent with Hyades membership. Triangles : candidates whose
   proper motion is consistent with Hyades membership (see text). The
   2 most promising substellar cluster candidates are shown by large
   triangles. NextGen (0.07-0.3~M$_\odot$), Dusty
   (0.04-0.07~M$_\odot$) and Cond (0.015-0.05~M$_\odot$) isochrones
   are shown and labelled with mass. In the (I, I-K) CMD, the dotted
   line indicates the locus of M8-T5 field dwarfs (from Dahn et
   al. 2002). The rms photometric error is shown as bars.  }
         \label{CMDiik}
   \end{figure*}
%
%______________________________________________________________

\subsection{Follow up K-band photometry}

Follow up K'-band imaging was obtained for 108 of the 125 optically
selected candidate members. About half of this sample was observed in
Nov.04 and the other half in Nov.05 using the 1k$\times$1k CFHT IR
camera (Starr et al. 2000). The exposure time varied from 0.5 to 14
minutes for candidates with I magnitudes ranging from 13 to 23. Each
object was dithered on 7 positions on the detector, so that the median
of the 7 images would provide an estimate of the sky background during
the exposure. Individual images were dome flat fielded, sky
subtracted, then registered and added to yield the final K'-band
image. K-band photometric standards from Hunt et al. (1998) were
observed every couple of hours each night. Aperture photometry was
performed on candidates and photometric standards. For 2 close visual
binary systems (CFHT-Hy-15/17, CFHT-Hy-4/4c), a large aperture was
first used to estimate the system's total flux, and PSF photometry was
performed to derive the flux ratio of the system components.

The derived K-band photometric zero point is 22.86 $\pm$ 0.03 mag,
assuming an average extinction coefficient of 0.07 mag/airmass. Slight
photometric variations occurred during some nights due to varying
extinction by thin cirrus. Images obtained on these nights were
recalibrated using 2MASS images of the same fields. In this case, the
photometric accuracy is of order of 0.05 mag. The other nights had
photometric conditions and the derived K-band magnitudes have an
accuracy of order of 0.02 mag. The (I, I-z) and (I, I-K) CMDs for the
108 candidates followed up in the K-band are shown in Fig~\ref{CMDiik}. 

\subsection{Astrometry and proper motion}

%Table~\ref{membership} : Hyades candidate membership
%name, RA, DEC, I, I-z, I-K, mu_a, mu_d,Membership, other name; 
%Footnotes : binaries (sep, PA)

%----------------------------------------------------------- Fig.5
% Fig.5 = /home/jbouvier/CFHTIR05B/PPM/CROSSCOR

   \begin{figure}[t]
   \centering
   \includegraphics[width=9cm]{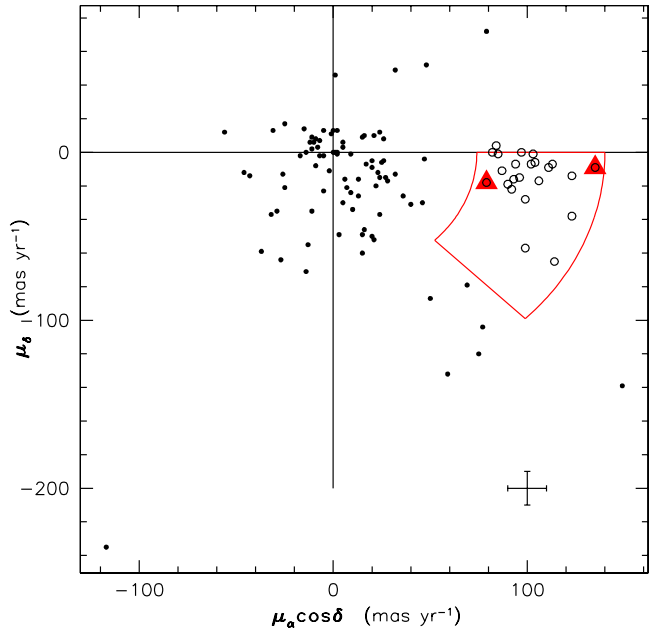}
      \caption{Proper motion vector diagram for 107 optically selected
        candidates followed up in the K-band (see text). The expected
        proper motion for Hyades members is shown by the (red) box
        (Bryja et al. 2004). Within these boundaries, 23 optically
        selected candidates (empty circles) are found to share the
        proper motion of the cluster, including 2 brown dwarfs (large
        triangles). Typical rms errors on the ppm measurements are
        shown by a cross. }
         \label{ppm}
   \end{figure}
%
%______________________________________________________________

The proper motion of optically selected Hyades candidates was computed
from pairs of optical (I, z) and infrared (K) images obtained 2 or 3
years apart (Nov.04, Nov.05) under similar seeing conditions
(0.6-0.8$\arcsec$ FWHM). The optical images were astrometrically
calibrated through CFHT's Elixir pipe-line (Magnier \& Cuillandre
2004) with an accuracy of 0.8$\arcsec$ rms. The infrared images were
astrometrically calibrated using the WCStools package (Mink 2006) and
2MASS images as references. We found the CFHT IR detector to be
aligned along the North-South direction to within 1 deg and the pixel
scale to be 0.211$\arcsec$, as mentionned in CFHT IR's Reference
Manual (0.211$\pm$0.001$\arcsec$). Some CFHT IR images had not enough
common objects with the shallower 2MASS images to allow a straight
astrometric calibration through this method. In this case, we adopted
the average orientation and plate scale derived for other CFHT IR
images and simply calibrated the location of the center of the field.

Pairs of optical and infrared images where then matched by celestial
coordinates to yield a list of common objects in the 2 images. The
(X,Y) location of the objects on both detectors was measured using
IRAF/DAOFIND package. A geometrical transformation was then computed
(IRAF/GEOTRANS) which maps the (X,Y) coordinates of the optical
detector into that of the IR detector. The geometrical transformation
was derived from a set of typically 10 to 20 reference stars evenly
distributed around the Hyades candidate. The candidate itself is not
included in the computation and an iterative procedure is applied to
reject any reference star exhibiting high residuals. The accuracy of
the geometrical transformation was typically 0.07 pixels rms.

The transformation was then applied to the Hyades candidate, mapping
its optical (X,Y) coordinates into the IR plane. The candidate proper
motion was derived from the difference between the observed and mapped
coordinates on the IR detector, divided by the time elapsed between
the 2 images, and multiplied by the pixel scale. For each candidate,
proper motion was measured using different sets of reference stars
within the image, and from both the I and z images. The various
measurements were always found to be consistent within errors and we
finally considered their weighted mean as the best estimate of the
candidate's proper motion.

\begin{table*}[t]
\caption{Probable Hyades members based on photometry and proper motion.}             % title of Table
\label{good}      % is used to refer this table in the text
\centering                          % used for centering table
\begin{tabular}{l l l l l l l l l l}        % centered columns (4 columns)
\hline\hline                 % inserts double horizontal lines
CFHT-Hy-\# &Internal & RA(2000) & Dec(2000) & I & I-z & I-K & $\mu_{\alpha\cos\delta}$& $\mu_\delta$ & Other name$^a$\\    % table heading 
&name&&&&&& \multicolumn{2}{l}{($mas.yr^{-1}$)}  \\
\hline                        % inserts single horizontal line
CFHT-Hy-1&HA10c00-33s  &  4 23 12.5  &  15 42 47  &  12.87  &  0.54  &  2.5  &  111  &  -9  & VA~262\\
CFHT-Hy-2&HC13c03-43s  &  4 27 24  &  14 7 6  &  12.91  &  0.46  &  2.41  &  123  &  -14  & VA~432\\
CFHT-Hy-3&HC06c00-511s  &  4 18 33.8  &  18 21 53  &  13.29  &  0.59  &  2.69  &  114  &  -65  & LP~414-158\\
CFHT-Hy-4$^1$&HC13c11-268s$^1$  &  4 28 22.4  &  13 49 22  &  13.35  &  0.57  &  2.61  &  84  &  4  & HAN~430\\
CFHT-Hy-5&HA2Ic06-310s  &  4 25 16.5  &  16 18 7  &  13.4  &  0.55  &  2.45  &  87  &  -11  & VA~352\\
CFHT-Hy-6&HB10c09-178s  &  4 24 31.5  &  13 55 41  &  13.48  &  0.52  &  2.41  &  97  &  0  & VA~326\\
CFHT-Hy-7&HB8Ic01-117s  &  4 20 56.1  &  14 51 33  &  13.53  &  0.52  &  2.5  &  106  &  -17  & VA~203\\
CFHT-Hy-8&HA11c00-112s  &  4 23 1.4  &  15 13 41  &  13.59  &  0.52  &  2.43  &  103  &  -1  & VA~260\\
CFHT-Hy-9&HB12c04-432s  &  4 30 42.5  &  14 39 41  &  13.59  &  0.62  &  2.59  &  113  &  -7  & HAN~495\\
(---$^2$&HB6Ic04-77s$^2$  &  4 22 32.8  &  15 51 24  &  13.67  &  0.55  &  2.32  &  92  &  -22  & VA~241)\\
CFHT-Hy-10$^3$&HC04c10-516s$^3$  &  4 23 56.7  &  18 38 19  &  14.17  &  0.6  &  2.68  &  102  &  -7  & RHy~191\\
CFHT-Hy-11&HC16c06-212s  &  4 34 21  &  14 51 16  &  14.39  &  0.57  &  2.65  &  82  &  0  & \\
CFHT-Hy-12&HB13c01-162s  &  4 31 16.3  &  15 0 12  &  14.44  &  0.64  &  2.7  &  94  &  -7  & \\
CFHT-Hy-13&HC01c00-279s  &  4 19 5.7  &  17 34 22  &  14.52  &  0.71  &  3.01  &  85  &  -1  & \\
CFHT-Hy-14&HA5Ic02-358s  &  4 30 4.2  &  16 4 6  &  14.71  &  0.69  &  2.76  &  99  &  -57  & RHy~281\\
CFHT-Hy-15$^4$&HA2Ic03-783s$^4$  &  4 27 6.4  &  16 25 48  &  14.78  &  0.71  &  2.76  &  93  &  -16  & RHy~240A\\
CFHT-Hy-16&HC19c03-841s  &  4 29 2.9  &  13 37 59  &  14.81  &  0.82&  3.19  &104  &  -6  & LH~91, LHD~0426+1331\\
CFHT-Hy-17$^4$&HA2Ic03-781s$^4$  &  4 27 6.6  &  16 25 46  &  14.83  &  0.71  &  2.76  &  96  &  -15  & RHy~240B\\
CFHT-Hy-18&HC09c09-231s  &  4 17 32.2  &  16 56 59  &  15.2  &  0.75  &  3.17  &  123  &  -38  & RHy~138\\
CFHT-Hy-19$^5$&HC08c07-331s$^5$  &  4 17 24.8  &  16 34 36  &  17.49  &  1.18  &  4.59  &  99  &  -28  & 2MASSW~J0417247+163436\\
CFHT-Hy-20$^6$&HC22c09-1735l$^6$  &  4 30 38.7  &  13 9 57  &  21.58  &  1.79  &  5.5  &  135  &  -9  & \\
CFHT-Hy-21$^7$&HA9Ic04-1346l$^7$  &  4 29 22.7  &  15 35 29  &  22.16  &  1.36  &  5.57  &  79  &  -18  & \\
\hline                                   %inserts single line
\multicolumn{10}{l}{$^a$ Ref. : VA = van Altena (1969); LP =
  Luyten et al. (1981); HAN = Hanson (1975); RHy = Reid (1992); LH =
  Leggett \& Hawkins (1988);}\\
\multicolumn{9}{l}{LHD = Leggett, Harris \& Dahn (1994);
  2MASS = Gizis et al. (1999)} \\
\multicolumn{10}{l}{$^1$ A faint companion is seen 3.2$\arcsec$ away
  from this Hyades probable member but its I-K color and proper motion indicates a background star.}\\
\multicolumn{10}{l}{$^2$ Classified as a non member by Reid
  (1993). Spectroscopically confirmed as a non-member in this study (see
  text.)}\\
\multicolumn{10}{l}{$^3$ Classified as a non member by Reid
  (1993). Spectroscopically confirmed as a member in this study (see
  text.)}\\
\multicolumn{10}{l}{$^4$ CFHT-Hy-15 and 17 form a
  visual pair (RHy~240 AB).}\\
\multicolumn{10}{l}{$^5$ Classified as a non member by Gizis et
  al. (1999), see text.}\\
\multicolumn{10}{l}{$^6$ This probable Hyades BD was also observed in
  the J and H bands, yielding (J-H, H-K)=(0.51, 0.43).}\\
\multicolumn{10}{l}{$^7$ This probable Hyades BD was also observed in
  the J and H bands, yielding (J-H, H-K)=(1.12, 0.77).}\\
\end{tabular}
\end{table*}

The results are listed as $\mu_{\alpha\cos\delta}$ and $\mu_\delta$ in
Tables~\ref{good} and \ref{wrong}. The rms accuracy on the ppm
measurement is usually of order of 10 mas/yr.  However, in some cases
when the distribution of reference stars is asymetric relative to the
candidate (i.e. when the optical candidate lies close to the CCD
edge), higher systematic errors might be expected. Also, one candidate
(HA13c08-1866l) was too faint in K (a probable non member) to have its proper
motion measured. The proper motion vector diagram of 107 optically
selected Hyades candidate members is shown in Figure~\ref{ppm}.

\subsection{Optical Spectroscopy}

We obtain optical spectroscopy for 3 new Hyades low mass candidate
members and for 2 other candidates recovered in our survey but
previously classified as non members (see Sect. 3.1). Spectroscopic
observations were performed from December 5 to 8, 2006 at ESO/NTT with
EMMI spectrograph which provides echelle spectra covering the
wavelength range from 580 to 990~nm at a resolution of 9,800. Using
NOAO/IRAF, the spectral orders were extracted, flat-field corrected,
wavelength calibrated using a Th-Ar lamp and corrected from the
instrumental response computed from spectrophotometric standards
observed during the same nights. Exposure times from 600 to 1200s
yielded a S/N ratio of about 30 per resolution element at 850 nm.

\subsection{Infrared Spectroscopy}

The three faintest probable Hyades members (Table~\ref{good}) were
observed with the NICS near infrared spectrograph at the 3.58m
Telescopio Nazionale Galileo (TNG). We employed the high efficiency,
very low resolution Amici spectroscopic mode, which yields a complete
0.85-2.45~$\mu$m spectra with a resolving power of 50 (Baffa et al.~
\cite{Bea01}; Oliva et al.~\cite{O03}). Observations were carried out
in service mode by the TNG staff on Oct. 14 and 18, 2005. We used the
1$^{\prime\prime}$ wide slit and total on-source integration times of
32, 56 and 40 minutes for CFHT-Hy-19, 20 and 21 respectively. The data
reduction procedure closely followed the one described in Testi et
al.~(2001). Correction for telluric absorption was achieved using
observations of the AS11 A0 star from the list of Hunt et al.~(\cite
{Hea98}) obtained during the same nights at a similar airmass as our
main targets and with an identical instrumental setup.

\section{Results}

We identified Hyades candidate members from their location in optical
and IR color-magnitude diagrams (CMDs) and assess their membership
status from proper motion measurements. We describe below the
selection criteria we used and applied to all stellar like objects
detected on the optical images.

\subsection{Probable Hyades members}

Possible Hyades members are selected from their location in the (I,
I-z) color magnitude diagram (CMD) relative to model isochrones. The
600 Myr NextGen (Baraffe et al. 1998) and Dusty (Chabrier et al. 2000)
isochrones computed for CFHT 12K I and z filters were used. In order to
account for photometric uncertainties and for the cluster's depth, the
selection boundary was taken to lie 0.1 mag bluer than the isochrones,
with the NextGen to Dusty transition occurring at I$\simeq$18.5 mag. The
(I, I-z) CMD is shown in Fig.~\ref{CMDiiz} together with the model
isochrones labelled with mass shifted at a distance modulus of 3.33
for the Hyades cluster (Perryman et al. 1998). The selection resulted
in 125 possible candidate members from I$\sim$13 to 23, covering the
mass range from 0.3 M$_\odot$ to about 45 Jupiter masses. The
stellar/substellar boundary occurs at I$\sim$17.8 in the CMD. Of the
125 possible candidates, 42 are low mass stars (0.1-0.3 M$_\odot$), 4
lie at the stellar/substellar boundary, and 79 lie in the substellar
range (M$\leq$0.072M$_\odot$). The list of optically selected
candidates is given in Tables~\ref{good} and \ref{wrong}.

Follow-up K-band imaging of 108 possible candidates provided
additional membership criteria. The (I, I-K) CMD of these candidates
is shown in Fig.~\ref{CMDiik}. Most of the low mass stellar candidates
qualify as possible members in this diagram as well, exhibiting (I-K)
colors consistent with model predictions. In contrast, a large
fraction of the optically selected substellar candidates appear too
blue compared to the Dusty model isochrones, which suggests they are
older late type field dwarfs unrelated to the cluster. However, the
model isochrones become increasingly uncertain at very low masses as
dust develops and may condense in the lower atmospheric layers, which
results in a sudden change of the I-K index towards much bluer
colors. The location of low mass substellar cluster members in the (I,
I-K) diagram may thus span a large range of colors, from the Dusty
isochrone bluewards. Chabrier et al. (2000) predict that the
transition between the Dusty and Cond models should occur at a
temperature of 1300-1400~K, which corresponds to a mass of 40 Jupiter
masses for Hyades members.

As an additional membership criterion, proper motion was computed for
all the candidates followed up in the K-band, except one
(HA13c08-1866l) which was too faint in K and therefore a probable non
member. The proper motion vector diagram of 107 optically selected
candidates is shown in Fig~\ref{ppm}. While most of the candidates
scatter near the origin, with relatively small proper motion and thus
most likely background field contaminants, 23 candidates have a proper
motion consistent with Hyades membership. The ppm vector diagram also
reveals 7 high proper motion objects ($\mu>$100 mas/yr), most likely
foreground late-type field dwarfs.

In order to analyse the distribution of objects in the ppm diagram, we
built a synthetic ppm vector diagram from simulated observations
spanning 1 sq.deg. in the direction of the Hyades cluster. The
simulated sample was obtained from the Besancon model of galactic
structure and kinematics (Robin et al. 2003). We extracted from this
sample all the objects lying in the same part of the (I, I-K) CMD as
the optically selected Hyades candidates. The ppm vector diagram of
the resulting synthetic subsample strongly ressembles the observed
one, with most objects scattered around the origin with a slightly
elongated distribution towards the south-east direction, and a few
high velocity foreground dwarfs. No object from the simulated sample
is found to lie within the boundary of the Hyades cluster proper
motion. This indicates that the contamination of the observed ppm
vector diagram by foreground field dwarfs at the expected location of
Hyades members is small. We find only one optically selected candidate
(HA12c07-101l) with a proper motion consistent with Hyades membership
which is too blue in the (I, I-K) CMD to be a member
(cf. Table~\ref{wrong}). We classified it as a non member.

%----------------------------------------------------------- Fig.5
% figure build in /home/jbouvier/HyadesABC/RadDist/12K

   \begin{figure}[t]
   \centering
   \includegraphics[width=9cm]{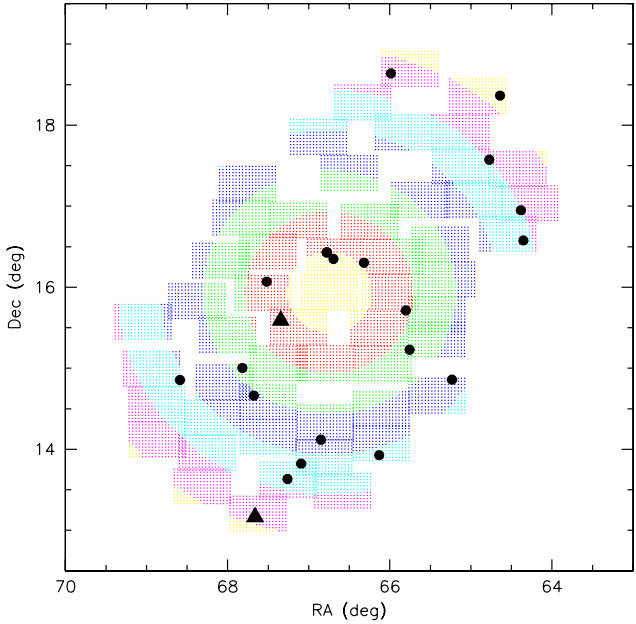}
      \caption{The spatial distribution of probable Hyades members
      reported from our survey. Low-mass stellar members are shown as
      large dots, and the 2 brown dwarfs by triangles. The separation
      of the 2 components of the visual binary CFHT-Hy-15/17
      (RHy~240AB) has been exagerated for clarity. The background
      shows the location of the 12K fields (the color code illustrates
      0.5~deg-wide rings around the cluster center. }
         \label{raddist}
   \end{figure}

%
%______________________________________________________________
%----------------------------------------------------------- Fig.5
% Fig.6 = /home/jbouvier/HyadesABC/NTT_EMMI/Spectra

   \begin{figure}[t]
   \centering
   \includegraphics[width=9cm]{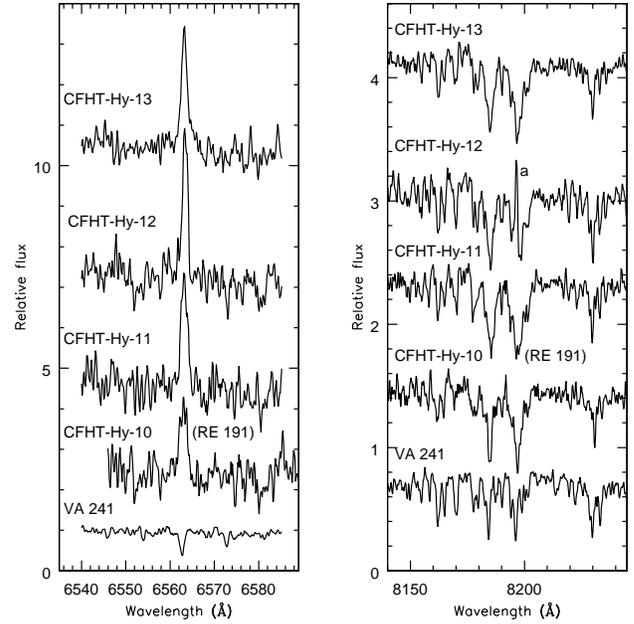}
      \caption{Optical spectra of 5 optically-selected Hyades very low
      mass stellar candidate members. The H$_\alpha$ region is shown
      in the left panel, the NaI doublet in the right panel. The NaI
      doublet 8195\AA\ line of CFHT-Hy-12 is affected by a bad pixel,
      labelled with ``a''. }
         \label{spectro}
   \end{figure}

%
%______________________________________________________________

Eventually, 22 candidates consistently qualify as probable Hyades
members on the basis of their optical photometry, their (I-K) color
and their proper motion. Their properties are listed in
Table~\ref{good} and their spatial distribution is shown in
Fig.~\ref{raddist}. Of these, 14 were already listed as probable
Hyades members in Prosser \& Stauffer's Open Cluster
Database\footnote{Open Cluster Database, as provided by
C.F. Prosser and J.R. Stauffer, and which currently may be accessed at
http://www.noao.edu/noao/staff/cprosser/, or by anonymous ftp to
140.252.1.11, cd /pub/prosser/clusters/.}
(P\&S database) based on various criteria and are confirmed
here. Their alternate names are given in Table~\ref{good}. Another 3
probable members in our survey had been previously reported but
classified as probable non members. One (CFHT-Hy-10 = RHy~191) is
listed in the P\&S database and classified as a probable non member
based on the lack of H$_\alpha$ emission in a low resolution spectrum
(Stauffer's priv. comm., quoted in Reid 1993). The higher resolution
spectrum we report for this object exhibits weak H$_\alpha$ emission,
with EW(H$_\alpha$)=3\AA, and deep NaI 8183,8195\AA\ lines, consistent
with membership (cf. Fig.~\ref{spectro}). Another candidate
(HB6Ic04$\_$77s = VA~241), also listed in the P\&S database, was
classified as a non member by Reid (1993) based on ``(V-I) colours
inconsistent with Hyades membership''. The high resolution spectrum we
obtain for VA~241 shows H$_\alpha$ in absorption and shallow NaI
doublet lines, a non member indeed (Fig.~\ref{spectro}). The last one
(CFHT-Hy-19) is located close to the stellar/substellar boundary and
has previously been reported by Gizis et al. (1999) from the 2MASS
survey of the Hyades cluster. This candidate is discussed in more
details in the next section, as the evidence for membership is
somewhat inconclusive. The remaining 5 probable members we report are
new. They include 3 low mass stars ($\sim$0.14~M$_\odot$) and 2
objects well within the substellar regime ($\sim$0.050~M$_\odot$). The
spectra of the 3 new low mass stars exhibit H$_\alpha$ emission with
equivalent widths between 4 and 6 \AA, and deep NaI doublet, fully
consistent with membership (cf. Fig.~\ref{spectro}).

In addition, we found during this survey an extremely red object,
HB6Ic09$\_$322l, with I=22.92 and I-K=7.55 ! The object appears point
like in the IzK images but is not a Hyades member according to its
proper motion (Table~\ref{wrong}). We note that this object lies only
3$\arcmin$ away from the Class 0 protostar IRAM 04191+1522 (Andr\'e et
al. 1999) and from the Class I protostar IRAS 04191+1523 (Lee et
al. 2002), both associated to the molecular condensations MC 18a and
18b (Onishi et al. 2002). This indicates that star formation is still
active in this area located at the outskirt of the Taurus molecular
cloud. HB6Ic09$\_$322l, whose proper motion is consistent with Taurus
membership (Frink et al. 1997), could thus be a new, self-embedded
Class I source. However, there are no 2MASS nor IRAS counterparts at
this location, nor is it detected by Spitzer at 24 and 70 $\mu$m
(Ph.~Andr\'e, priv. comm.).

\subsection{Comparison with previous Hyades surveys}

In order to check the consistency of our membership classification of
the 125 optically selected candidates, we searched the literature to
compare our results with those of previous deep surveys of the Hyades
cluster.

%----------------------------------------------------------- Tab.4
%\begin{table}
%\caption{Comparison between Dobbie et al.'s (2002) measurements and ours
%  for common Hyades candidates.}             % title of Table
%\label{dobbie}      % is used to refer this table in the text
%\centering                          % used for centering table
%\begin{tabular}{l l l l l l}        % centered columns (4 columns)
%\hline\hline                 % inserts double horizontal lines
%Name & I & z & K & $\mu_{\alpha\cos\delta}$ & $\mu_\delta$ \\    % table heading 
%&&&& ($mas.yr^{-1}$) & ($mas.yr^{-1}$) \\
%\hline                        % inserts single horizontal line
%HB7Ic04$\_$145s  & 15.36   &     14.58 &   11.32  &    25    &    -6\\
%=h4046b    &    15.85    &    14.51  &  11.24  &    13$\pm$18   &  -8$\pm$16\\
%\ \\
%HB7Ic11$\_$463s &  19.39    &    18.30 &   15.20  &    40    &    -31\\
%=h7334b    &    19.72   &     18.25  &  ----    &   13$\pm$18   &  -61$\pm$21\\
%\hline                                   %inserts single line
%\end{tabular}
%\end{table}

Dobbie et al. (2002) conducted a wide (10.5 sq. deg.) optical (I,z)
survey of the Hyades cluster down to a completeness limit of
I$\sim$20.3 mag. They report 20 possible candidate members, of which
they confirm only one as being a (previously known) proper motion low
mass member (h5078e = RHy~297). Of the 20 candidates they report, 8
fall in the area we surveyed. Of these, 4 are too blue in the (I, I-z)
CMD to be selected as possible members (h8644b, h1272b, h8711b,
h5044b), and 2 are saturated on our images (h9448b, h4239b). The 2
remaining ones are recovered here as possible members in the (I, I-z)
CMD but are eventually rejected on the basis of their discrepant
proper motion (HB7Ic04$\_$145s = h4046b; HB7Ic11$\_$463s =
h7334b). 

%Table~\ref{dobbie} shows a comparison between Dobbie et al.'s
%measurements and ours for the 2 optically selected candidates common
%to the 2 surveys. A good agreement is found for z- and K-band
%photometry, as well as for proper motion within errors. However, there
%is a significant difference in the measured I-band magnitudes, Dobbie
%et al's being larger than ours by 0.3-0.5 mag, in spite of being
%brought to the same (Cousins) system. We note that Dobbie et
%al. (2002) applied color term corrections to their I-band photometry,
%while we did not as the Lyon model isochrones were computed for CFHT
%filters, and this might be the origin of these discrepancies. In any
%case, since we recover previously known Hyades members and identify
%new ones which all lie within 0.1 mag from the 600 Myr NextGen
%isochrone in the (I, I-z) CMD (see Fig~\ref{CMDiik}), this gives us
%confidence that our photometry is not affected by large systematic
%errors.

Reid \& Hawley (1999, RH99) reported low resolution follow up
spectroscopy of 12 possible low mass candidate members identified by
Leggett \& Hawkins (1989, LH) from a Schmidt survey of the central
regions of the cluster. Based on radial velocity measurements, they
find that only one of the 12 stars is a probable low mass cluster
member (LH~0418+13). Of the 12 LH objects, 6 fall in the cluster area
we surveyed. Of these, 4 are too blue in our (I, I-z) CMD to be
selected as possible cluster members (LH~0427+12, 0418+15, 0419+15,
0420+15). Another one (LH~0422+17 = HB1Ic01$\_$386l) was selected as a
possible member from its location in the (I, I-z) CMD but its proper
motion proved inconsistent with membership. RH99 also measured a
discrepant radial velocity for this object, which appears to be an
unrelated field dwarf. The last object (LH~0424+15) falls on a bad
column of a CCD preventing us from deriving its photometry.

Gizis et al. (1999, GRM99) searched for very low mass stars and brown
dwarfs in the central region of the Hyades cluster using the 2MASS
survey. They selected 40 possible candidates, including 29 new ones,
from their infrared colors corresponding to spectral types M8-L4. From
the follow up low resolution spectroscopy they performed, none of
these candidates appear to be cluster members. Eleven GRM99 new
candidates fall in the area covered by our survey. Of these 11, 10 are
too blue in the (I, I-z) CMD to be selected as possible cluster
members (2MASSW~J0434462+144802, J0435489+153719, J0431500+152814,
J0431322+152620, J0431420+162232, J0421175+153003, J0421521+151941,
J0424045+143129, J0430232+151436, J0423242+155954). The last one,
however, CFHT-Hy-19 = 2MASSW~J0417247+163436, lies close to the
stellar/substellar boundary in Fig.~\ref{CMDiiz}. We classify it as a
probable cluster member based on its optical photometry, (I-K) color
and proper motion, all being consistent with membership. Gizis et
al. (1999) dismissed this candidate as a cluster member based on its
low resolution spectrum (Sp.T. M8) exhibiting weak NaI 8200\AA\
equivalent width, suggestive of low surface gravity. They also fail to
detect H$_\alpha$ in emission in their low resolution
spectrum. Comparing with theoretical models, they estimate a surface
gravity of $\log g$=4.4--4.7, somewhat lower than expected for Hyades
members ($\log g$$\sim$5.0).  They conclude it is presumably a young
(10-30 Myr), background (d$\sim$150-200 pc) substellar object
(M$\leq$0.04$M_\odot$), possibly associated with the nearby Taurus
star forming region. We find this possibility unlikely however, as the
proper motion we measure for this candidate ([99, -28] mas/yr) is much
larger than the proper motion of Taurus SFR members (typically,
[10,-20] mas/yr, Frink et al. 1997).  Furthermore, the candidate's
proper motion is fully consistent with Hyades membership. Hence,
pending additional evidence for (or against) membership, we retain
this candidate as a probable member.

Finally, Prosser \& Stauffer's Open Cluster Database lists 147
probable members with V$\geq$15.3 (I$\geq$12.5). Of these, 30 lie in
the area covered by our survey: 14 are recovered and confirmed here as
members (see Table~\ref{good}, VA~262, VA~432, LP~414-158, HAN~430, VA~352,
VA~203, VA~326, VA~260, HAN~495, RHy~281, RHy~138, RHy~240A, RHy~240B, LH~91); 9
others are saturated on our short exposure images, with I magnitudes
in the range from 12.52 to 13.36 and are therefore not considered in
our study (VA~329, VA~213, VA~216, VA~94, VA~362, LP~415-875, VA~763, VA~127,
VA~368, all of which except VA~329 have been spectroscopically confirmed
as members, e.g., Reid \& Mahoney 2000); another one is HAN~150 (VA~95,
RHy~133, HC08c00$\_$261s) which we do recover in our survey but has too
blue an (I-z) color to be a candidate member (I=15.72, I-z=0.4). It
was already rejected by Reid et al. (1993) based on its too blue (V-I)
color. The 6 remaining objects are the RH99 candidates discussed
above, none of which qualify as a member in our study. Hence, we
recover all the candidate probable members listed in the P\&S database
which lie in the area covered by our survey and provide a new
membership assessment for most of them.  Another 6 objects listed as
probable members in the P\&S database lie in the spatial limit of our
survey but fall in gaps between our fields, as we avoided the
immediate vicinity of bright stars. These are RHy~159, RHy~228, RHy~230,
RHy~287 (Reid 1992) and LH~0416+16, LH~0429+15 (Legget \& Hawkins
1988). Two of these (RHy~228 and RHy~230) have been confirmed as members
(Reid et al. 1995; Reid \& Mahoney 2000), 2 others (RHy~159, RHy~287) have
not been followed up (Reid 1992), and the 2 remaining ones have been
shown to be non-members (Reid \& Hawley 1999). Conversely, as
discussed above (see Section 3.1), we recovered 2 objects, previously
classified as non-members in the P\&S database, one which we confirm
as being a non-member (VA~241), the other which we classify as a {\it
bona fide} Hyades member (CFHT-Hy-10 = RHy~191).

\section{Discussion}

\subsection{L and T dwarfs in the Hyades cluster}

We report here the first 2 Hyades brown dwarf candidates (CFHT-Hy-20,
21) as well as a previously detected very low mass object (CFHT-Hy-19)
close to the stellar-substellar boundary. The 3 objects are strong
candidate cluster members as their optical/infrared luminosity and
colors are consistent with membership, as is their proper motion. The
2 BDs are well within the substellar domain with an estimated mass of
about 50 Jupiter masses (see Table~\ref{mass}) while the lowest mass star
has an estimated mass around 0.08~M$_\odot$. A finding chart for the 2
substellar objects is provided in Fig.~\ref{fc}. According to models,
at an age of 625~Myr, these objects should have an effective
temperature around 1600~K for the 2 BDs and around 2400~K for the VLM
star, which corresponds to spectral types late-L/early-T and late-M
respectively.

\begin{figure}[t]
   \centering
   \includegraphics[width=9cm]{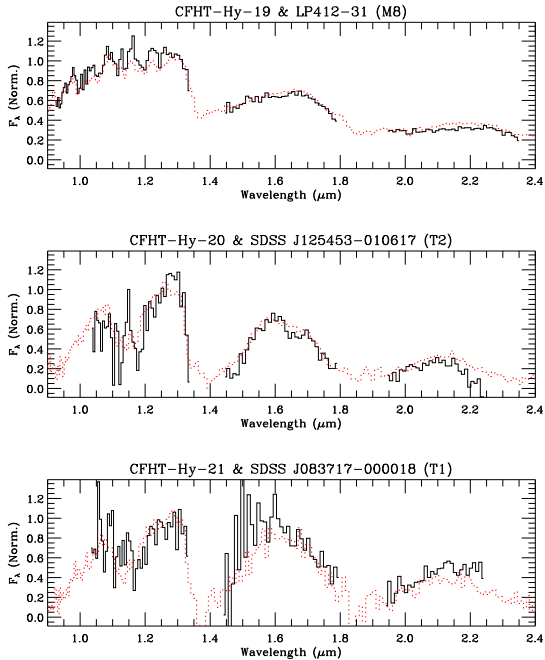}
   \caption{Near-infrared Amici low resolution spectra of CFHT-Hy-19,
20 and 21 (solid lines from top to bottom). In each panel we also show
the closest matching field dwarf spectrum from the low resolution
Amici spectral library (Testi et al.~\cite {Tea01};
Testi~\cite{T04}).}
   \label{nics}
   \end{figure}

The TNG/Amici low resolution spectra we obtained for these 3 objects,
CFHT-Hy-19, 20 and 21 are shown in Figure~\ref{nics}. The portions of
the spectra that are affected by the strongest telluric absorption
(around 1.4 and 1.9~$\mu$m) are not shown; the portion beyond
2.25$\mu$m and shortward of 1$\mu$m for CFHT-Hy-21 and CFHT-Hy-20 are also
not shown as the signal to noise ratio in the blue end of the spectra
is too low and, with the integration time chosen for these objects,
the red edge of the spectra is in a non linear regime of the detector
due to the high sky background.

The spectra of our targets have been compared with the Amici spectral
library of field dwarfs compiled by Testi et al.~(\cite{Tea01}) and
Testi~(\cite{T04}). While the gravity of field dwarfs is expected to
be somewhat different from that of Hyades members, at the Amici
resolution, the general shape of the spectrum should closely match for
similar spectral types. Following this expectation, we have matched
our observed spectra with the most similar field dwarf spectra and
derived an approximate spectral type for our targets. In
Figure~\ref{nics} we show as dotted lines the spectra of the closest
matching dwarfs together with our target stars. We find an excellent
match with dwarfs with spectral types M8, T2 and T1 for CFHT-Hy-19, 20
and 21 respectively; neglecting the possible systematic offset
introduced by comparing dwarfs of different gravity, the uncertainty
of our procedure for spectral typing is within one subclass. It is
worth pointing out that our near infrared classification of CFHT-Hy-19
agrees with the optical M8 classification of Gizis et al.~(1999).

%----------------------------------------------------------- 
% Fig.7 = from Leonardo (/home/jbouvier/HyadesABC/leonardo)

%______________________________________________________________

As noted above, the 2 T-dwarfs we report here are strong candidate
Hyades members with their photometry and proper motion consistent with
membership. Nevertheless, we proceed in estimating the probability
that they could be unrelated field T dwarfs projected onto the Hyades
cluster. The 2 observed T-dwarfs have a K-band magnitude of 16.08 and
16.59, respectively. T0-T2 field dwarfs at an age of a few Gyr have an
absolute K-band magnitude of $\simeq$13.5 (Chiu et al. 2006). This
would then locate the contaminating dwarfs at a distance between
approximately 30 and 40 pc. Combined with the area covered by our
survey, this corresponds to a volume of 65~pc$^3$. From the
combination of the 2MASS and SDSS DR1 surveys, Metchev et al. (2007)
derived an upper limit of 0.9$\times$10$^{-3}$ pc$^{-3}$ on the space
density of T0-T2.5 dwarfs in the solar neighborhood. Hence, we would
expect $\leq$0.06 early field T dwarf to contaminate our survey.
%Cruz
%et al. (2007) derived a space density of $\simeq$10$^{-3}$
%pc$^{-3}$mag$^{-1}$ for late L-dwarfs in the solar neighborhood and we
%take this estimate as an upper limit for the space density of early T
%dwarfs. We thus expect to detect at most 0.033 contaminating early
%field T dwarf in our survey in the range K=16.1-16.6 mag. 
An independent empirical estimate is provided by the detection rate of
early T dwarfs in the UKIDSS LAS survey. Chiu et al. (2007) reported
the detection of 3 early T-dwarfs up to a distance of 98~pc over
136~deg$^2$ in the southern sky. Scaling from their volume to ours, we
derive that about 0.06 early field T-dwarfs should contaminate our
survey. This further strengthens the likelihood that the 2 candidates
we report here are indeed the first brown dwarfs and the lowest mass
members of the Hyades cluster known to date.

\subsection{The Hyades mass function}

%
%______________________________________________________________
%----------------------------------------------------------- Fig.5
% Fig.8 = /home/jbouvier/HyadesABC/RadDist/MF/MF_final

   \begin{figure*}[t]
   \centering
   \includegraphics[width=13cm]{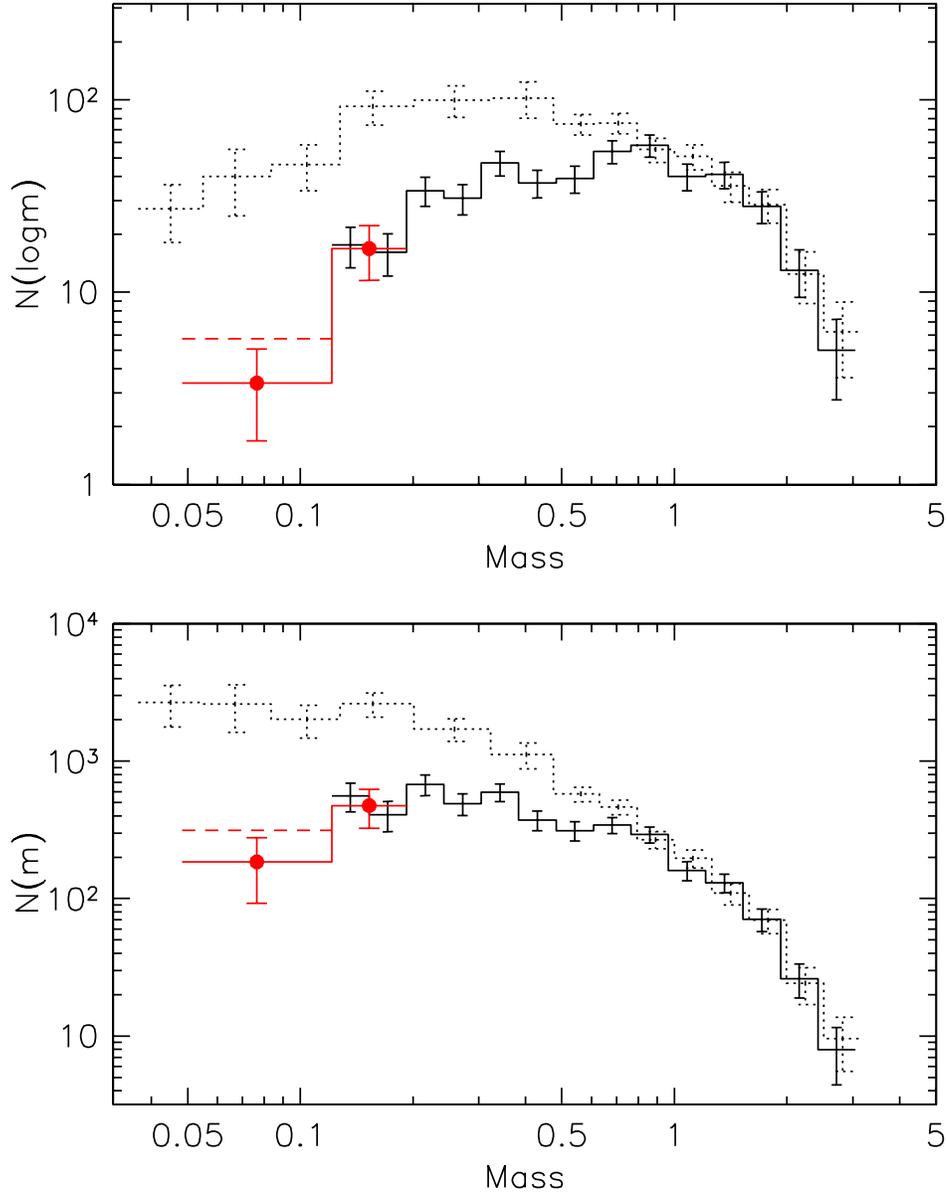}
      \caption{The mass distribution of Hyades probable members. {\it
      Upper panel~:} number of systems per 0.1 logarithmic mass
      bin. {\it Lower panel~:} number of systems per unit mass
      bin. {\it Solid histogram~:} mass distribution of Hyades
      systems. The mass distribution derived from our survey is shown
      as a red histogram in the 2 lowest mass bins
      (0.048-0.19~M$_\odot$). It has been normalized to the mass
      distribution of the P\&S database over the mass range
      0.12-0.19~M$_\odot$ (see text). The dashed histogram in the
      lowest mass bin (0.048-0.12~M$_\odot$) assumes a core radius of
      7~pc for brown dwarfs in the Hyades cluster (see text). {\it
      Dotted histogram~:} mass distribution of Pleiades systems from
      Moraux et al. (2003), normalized to the Hyades PDMF in the mass
      range 1-3~M$_\odot$. }
         \label{mf}
   \end{figure*}
%-----------------------------------------------------------------

Our survey is complete in the mass range from less than 50 Jupiter
masses up to 0.20~$M_\odot$. We are thus able to derive the cluster's
mass function across the stellar-substellar boundary and well into the
brown dwarf regime. The mass of the probable members we report in
Table~\ref{good} was derived using the 600~Myr NextGen (Baraffe et
al. 1998) and Dusty (Chabrier et al. 2000) model isochones, with the
transition between NextGen and Dusty taken at 0.08~M$_\odot$
(T$_{eff}$=2500K). These models provide Mass-Magnitude relationships
in the CFHT I and z system, as well as for the
K-band. Table~\ref{mass} lists the mass of the candidates derived from
their I, z and K magnitudes (DM=3.33 mag). Over the mass range
0.05-0.25~M$_\odot$, mass estimates obtained from the I and z
magnitudes agree to within 4\% with no systematic offsets. In
contrast, masses derived from K magnitude are usually higher than
those derived from the I magnitude by up to 20\% in the stellar
domain, and are lower by up to the same amount in the substellar
domain. The discrepancy between masses obtained from either visual or
near-IR magnitudes presumably results at least in part from model
uncertainties but might in some cases be related to the presence of
unresolved companions. These uncertainties do not significantly affect
the shape of the mass function we derive below and we adopt the masses
obtained from the I-band magnitude in the following.

%Table : 12K candidate mass = /home/jbouvier/HyadesABC/RadDist/MF/

\begin{table}[t]
\caption{Mass estimates for the 12K Hyades members.}             % title of Table
\label{mass}      % is used to refer this table in the text
\centering                          % used for centering table
\begin{tabular}{l l l l l l l}        % centered columns (4 columns)
\hline\hline                 % inserts double horizontal lines
CFHT-  &  I  &  Mass(I)  &  z  &  Mass(z)  &  K  &  Mass(K) \\
Hy-\# &&(M$_\odot$) && (M$_\odot$) && (M$_\odot$) \\
\hline                        % inserts single horizontal line
1  &  12.87  &  0.256  &  12.33  &  0.258  &  10.37  &  0.292 \\
2  &  12.91  &  0.251  &  12.45  &  0.244  &  10.5  &  0.273 \\
3  &  13.29  &  0.214  &  12.7  &  0.217  &  10.6  &  0.258 \\
4  &  13.35  &  0.208  &  12.78  &  0.209  &  10.74  &  0.239 \\
5  &  13.4  &  0.203  &  12.85  &  0.201  &  10.95  &  0.214 \\
6 &  13.48  &  0.196  &  12.96  &  0.192  &  11.07  &  0.2 \\
7  &  13.53  &  0.192  &  13.01  &  0.188  &  11.03  &  0.205 \\
8  &  13.59  &  0.188  &  13.07  &  0.183  &  11.16  &  0.192 \\
9  &  13.59  &  0.188  &  12.97  &  0.191  &  11  &  0.208 \\
10  &  14.17  &  0.151  &  13.57  &  0.150  &  11.49  &  0.164 \\
11  &  14.39  &  0.141  &  13.82  &  0.138  &  11.74  &  0.146 \\
12  &  14.44  &  0.139  &  13.8  &  0.139  &  11.74  &  0.146 \\
13  &  14.52  &  0.136  &  13.81  &  0.138  &  11.51  &  0.162 \\
14  &  14.71  &  0.128  &  14.02  &  0.129  &  11.95  &  0.133 \\
15  &  14.78  &  0.126  &  14.07  &  0.127  &  12.02  &  0.129 \\
16   &  14.81  &  0.125  &  13.99  &  0.130  &  11.62  &  0.154 \\
17 &  14.83  &  0.125  &  14.12  &  0.125  &  12.07  &  0.126 \\
18   &  15.2  &  0.113  &  14.45  &  0.113  &  12.03  &  0.128 \\
19  &  17.49  &  0.074  &  16.31  &  0.077  &  12.9  &  0.094 \\
20   &  21.58  &  0.050  &  19.79  &  0.052  &  16.08  &  0.039 \\
21   &  22.16  &  0.048  &  20.8  &  0.049  &  16.59  &  0.035 \\
\hline                                   %inserts single line
\end{tabular}
\end{table}

In order to build the complete present-day mass function (PDMF) of the
Hyades cluster, we combine the results of our survey in the
0.05-0.20~M$_\odot$ mass range to the P\&S database which covers the
mass range from about 0.1~$M_\odot$ to the most massive Hyades members
over the whole cluster area. Since we report only 3 new members (over
a total of 18, see Table~\ref{good} and~\ref{mass}) in the mass range
from 0.1 to 0.25~M$_\odot$ that were not previously listed in the P\&S
database, we assume that the database is nearly complete down to
0.1~M$_\odot$ (Reid 1993). Conversely, among the 30 low mass probable
members listed in the P\&S database that lie in the area of our
survey, 8 are rejected as being non members (see Section 3.2). We thus
estimate the contamination of the P\&S database by field dwarfs to be
about 27\% below 0.3~M$_\odot$. We assume the P\&S database to be
uncontaminated for brighter, better characterized Hyades members. The
mass of the probable members from the P\&S database was estimated from
their V-band magnitude as explained in Appendix~C.

Figure~\ref{mf} show the resulting mass distribution of Hyades
probable members over the whole mass range from 0.050 to 3~M$_\odot$,
obtained by combining the results of our survey and the P\&S
database. Note that since sub-arcsecond binaries are unresolved in
seeing-limited imaging surveys (Patience et al. 1997; Reid \& Gizis
1997), this mass distribution represents the systems mass
function. The mass function was built by counting the number of
objects in each mass bin and statistically correcting for the
contamination of the P\&S database by low mass field dwarfs. In order
to account for the different areas covered by the 2 datasets, the
number of objects detected in our survey was normalized to the number
of objects listed in the P\&S database over the common mass range
0.12-0.19~$M_\odot$ where the 2 datasets are complete. The
normalization factor is 3.4 with, respectively, 10 and 34 objects in
our survey and the P\&S database over this mass range. This correction
factor is consistent with the extrapolation of the areal coverage of
our survey to the whole cluster's area assuming the radial
distribution of low mass stars follows a King profile with a core
radius of 3~pc and a tidal radius of 10.3~pc (Perryman et al. 1998).

The estimate of the cluster's lower mass function we derive by simply
counting the number of BDs and VLM stars detected in our survey and
renormalizing to the whole cluster's area implicitely assumes that the
radial distribution of the two populations is the same. Alternatively,
if the Hyades cluster is fully relaxed, one may expect the core radius
to scale with the inverse square root of mass, i.e., that the BD
population are more widely distributed than the low mass stars. A core
radius of 3~pc for low mass stars would then translate into
$r_c\simeq$7 pc for brown dwarfs. Correcting for the different radial
distributions of BDs and low mass stars within the 10.3~pc cluster's
tidal radius, the actual number of brown dwarfs over the cluster's
area would then be about 70\% higher than the above estimate. Whenever
relevant, this additional correction is taken into account in the
discussion below.

Fig.~\ref{mf} compares the resulting Hyades PDMF to that of the
Pleiades over the mass range from 0.050 to 3~M$_\odot$. When expressed
as the number of objects per linear mass bin, the mass distribution
can be described as a power law over a limited mass range~: $\Psi(m) =
dN/dm\propto$ m$^{-\alpha}$ (Salpeter 1955; Kroupa 2002). In the
high mass range (M$\geq$1$M_\odot$), the Hyades and Pleiades mass
functions are found to be similar in shape, with a power law exponent
which agrees with a Salpeter slope ($\alpha$=2.35, Salpeter
1955). However, in the lower mass range, the mass functions of the 2
clusters clearly differ (cf. Fig.~\ref{mf})~: while the slope of the
power-law MF over the mass range 0.050-0.2~M$_\odot$ amounts to
$\alpha=$+0.6 for the Pleiades (Moraux et al. 2003), we find
$\alpha=$-1.3 for the Hyades over the same mass range (or
$\alpha=$-0.8 assuming a 7~pc core radius for Hyades BDs). From the
lack of brown dwarfs detection in previous surveys, Gizis et
al. (1999) found $\alpha=0$ to be an upper limit of the Hyades PDMF
below 0.3~M$_\odot$, which is consistent with our estimate. This points
to a strong deficiency of very low mass stars and substellar objects
relative to massive stars in the Hyades cluster compared to the MF of
Pleiades cluster.

%______________________________________________________________
% Fig.9 /home/jbouvier/HyadesABC/RadDist/MF/MF_final
   \begin{figure}
   \centering
   \includegraphics[width=8cm]{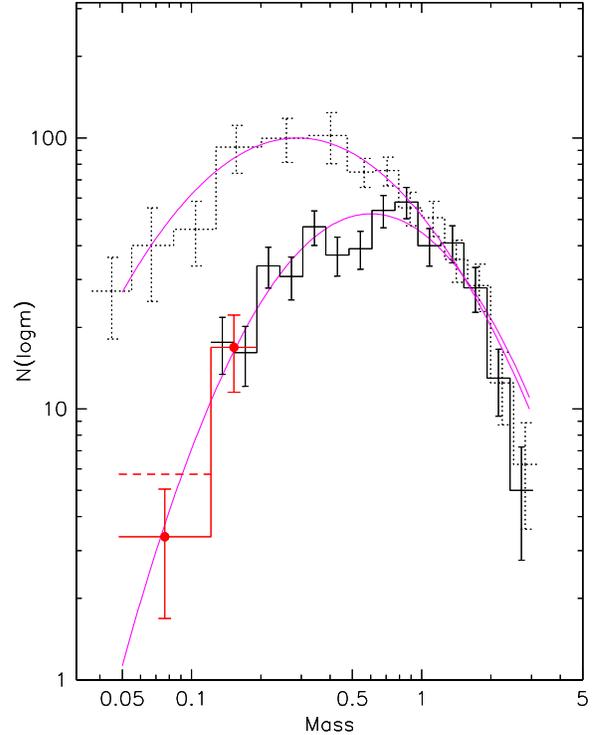}
      \caption{Lognormal fits to the mass distribution of the Hyades
      and Pleiades clusters over the mass range
      0.050-3.0~M$_\odot$. The observed mass distributions have been
      least-square fitted with a lognormal functional form~: $\xi(\log
      m) = k \cdot \exp\left[ - \frac{(\log m - \log m_0)^2}
      {2\sigma^{2}}\right]$ (see text). For the Hyades, we find
      $k$=52.4, $m_0$=0.60~M$_\odot$, and $\sigma$=0.39, while the
      best fit to the Pleiades yields $k$=100,
      $m_0$=0.29~M$_\odot$, and $\sigma$=0.47. }
         \label{mffit}
   \end{figure}
%------------------------------------------------------------------

When expressed as the number of objects per logarithmic mass bin, the
MF can be approximated by a lognormal functional form~: 
\begin{equation}
%  \label{eq:4}
  \xi(\log m) = dN/d(\log m) \propto \exp\left[ - \frac{(\log m - \log 
    m_0)^2} {2\sigma^{2}}\right]
\end{equation}
where $m_0$ is the peak mass and $\sigma$ the width of the
distribution (Miller \& Scalo 1979; Chabrier 2003). A least square fit
to the Hyades PDMF over the mass range 0.050 to 3~M$_\odot$ yields
$m_0\simeq0.60$~M$_\odot$ and $\sigma=0.39$ (or
$m_0\simeq0.59$~M$_\odot$ and $\sigma=0.40$ assuming a 7~pc core
radius for BDs). For the Pleiades it yields $m_0\simeq0.29$M$_\odot$
and $\sigma=0.47$. The lognormal fits to the mass functions of the 2
clusters are shown in Figure~\ref{mffit}.  The deficiency of low mass
members in the Hyades cluster's PDMF compared to the MF of the Pleiades
cluster is also clearly apparent from these lognormal representations.

Is the present-day Hyades mass function, as we observe it, a
dynamically evolved version of the Pleiades mass function ? A number
of N-body simulations have investigated the dynamical evolution of
young stellar clusters and allow us to address this issue. First, our
results indicate that the peak stellar mass in the Hyades cluster is
0.6~$M_\odot$, while it is only 0.3~M$_\odot$ in the Pleiades
cluster. This result is in qualitative agreement with Kroupa's (1995)
N-body models which predict that the mean cluster mass, at least in
the inner few parsecs of the system, linearily increases with
time. The increasing peak mass of aging clusters results from cluster
members experiencing weak gravitational interactions over a timescale
of several 100~Myr thus leading to mass segregation and the
preferential evaporation of low mass members (Terlevich 1987).

Numerical models of the dynamical evolution of young open clusters by
Adams et al. (2002) further predict that by an age of 625~Myr the
Hyades cluster has lost about 2/3 of its initial stellar population
and up to 70-90\% of its initial substellar population, depending on
the model assumptions. In contrast, at an age of only 120~Myr, the
Pleiades cluster is expected to have lost only 10-15\% of its initial
stellar and substellar members (see also de la Fuente Marcos \& de la
Fuente Marcos 2000). Assuming that the Hyades cluster is a
dynamically-evolved version of the Pleiades cluster, we can estimate
from the present-day MF of the 2 clusters the amount of low mass
members that have been lost over a timescale of a few 100~Myr on the
main sequence. Using the lognormal fits to the Hyades and Pleiades MF
obtained above, we find that the Hyades must have lost about 60\% of
its initial low mass stellar population (0.08-1.0~M$_\odot$) and about
95\% of its initial substellar population (or 92\% assuming a 7~pc
core radius for BDs in the Hyades). Within modelling and observational
uncertainties, these empirical estimates of the preferential
evaporation of low mass cluster members compare reasonably well with
model predictions.

Hence, the Hyades cluster may well be a dynamically-evolved version of
the Pleiades cluster indeed, as suggested by Kroupa (1995), i.e., both
clusters may have had similar initial mass functions down to the
substellar domain prior dynamical evolution, as also observed in other
young open clusters (de Wit et al. 2006; Moraux et al. 2007). This
would suggest that the IMF is not very sensitive to metallicity
either, as the Hyades is slightly more metal rich than other young
open clusters. Finally, by extrapolating the estimate of the
present-day Hyades mass function derived above to the lowest mass BDs
down to 0.013~M$_\odot$ and to the whole cluster area, we predict that
the Hyades cluster currently harbors $\sim$10-15 brown dwarfs while
it may have counted up to $\sim$150-200 such objects initially.

With a typical velocity of $\simeq$1~km~s$^{-1}$ (de la Fuente Marcos
\& de la Fuente Marcos 2000), a fraction of the substellar escapers
should now populate the solar neighborhood and be members of the
so-called Hyades moving group or ``supercluster'' (Eggen 1958, 1993;
Chereul, Cr\'ez\'e \& Bienaym\'e 1998; Chumak, Rastorguev \& Aarseth
2005). The space velocity of a few field brown dwarfs detected within
20~pc from the Sun indeed seem consistent with their being part of this
kinematical stream (Bannister \& Jameson 2007; Zapatero Osorio et
al. 2007).

%Cluster depth changes DM (see Perrymann et al.) : run MC simulations ?

\section{Conclusion}

From a deep, wide-field survey of the central region of the Hyades
cluster, we identified new very low-mass cluster members, including 2
T-dwarfs with a mass of $\simeq$ 50 Jupiter masses. The comparison of
the Hyades lower mass function with that of younger clusters indicates
that the cluster is strongly depleted in very low mass objects at an
age of 625~Myr. We thus estimate a total of about 15 brown dwarfs in
the present-day Hyades cluster compared to about 500 stars. This
depletion appears to result from the preferential evaporation of the
lowest mass cluster members over a timescale of several 100~Myr, as
predicted by N-body models of the dynamical evolution of young
clusters. A fraction of the substellar escapers should populate the
solar neighborhood, and may be detected as members of the Hyades
moving group from large scale surveys.

\begin{acknowledgements}
We thank the TNG staff for the support in preparing and executing the
service observing observations. This research benefited from financial
assistance from the European Union Research Training Network ``The
Formation and Evolution of Young Stellar Clusters''
(RTN1-1999-00436). GM acknowledges financial support by the DFG under
Grant ME2061/3-1. This research has made use of the SIMBAD database,
operated at CDS, Strasbourg, France, and of the IRAF package
distributed by the NOAO.
\end{acknowledgements}

\begin{appendix}

\section{Coordinates of CFHT12K Hyades fields}

%----------------------------------------------------------- Tab.1

\begin{table}[h]
\caption{CFHT 12K fields. Date of observation, coordinates of mosaic center,
  seeing FWHM.}             % title of Table
\label{fieldscoo}      % is used to refer this table in the text
\centering                          % used for centering table
\begin{tabular}{l l l l l}        % centered columns (4 columns)
\hline\hline                 % inserts double horizontal lines
Field & Date & RA & Dec. & Seeing\\    % table heading 
&&(2000)&(2000)& $\arcsec$ \\
\hline                        % inserts single horizontal line
HYADES-A1 & 2002-10-01 & 4:26:59.27 & 15:56:02.3 & 0.8 \\
HYADES-A2 & 2002-10-01 & 4:26:40.56 & 16:22:44.0 & 0.7 \\
HYADES-A3 & 2002-10-01 & 4:23:45.80 & 16:22:33.0 & 0.7 \\
HYADES-A4 & 2002-10-01 & 4:23:58.76 & 15:56:41.9 & 0.6 \\
HYADES-A5 & 2002-10-01 & 4:30:15.06 & 15:55:45.5 & 0.6 \\
HYADES-A6 & 2002-10-03 & 4:30:01.01 & 16:26:36.0 & 0.8 \\
HYADES-A7 & 2002-10-03 & 4:28:30.80 & 16:49:23.1 & 0.6 \\
HYADES-A8 & 2002-10-03 & 4:25:00.50 & 16:48:19.9 & 0.6 \\
HYADES-A9 & 2002-10-03 & 4:28:30.40 & 15:30:45.2 & 0.6 \\
HYADES-A10 & 2002-11-05 & 4:24:31.74 & 15:29:27.0 & 0.7 \\
HYADES-A11 & 2002-10-02 & 4:24:20.31 & 15:03:30.2 & 0.8 \\
HYADES-A12 & 2002-11-05 & 4:27:11.04 & 15:03:45.6 & 0.8 \\
HYADES-A13 & 2002-11-05 & 4:29:30.52 & 15:03:38.6 & 0.8 \\
HYADES-B1 & 2002-11-05 & 4:26:00.47 & 17:24:35.5 & 0.8 \\
HYADES-B2 & 2002-11-06 & 4:27:31.22 & 17:51:16.7 & 0.7 \\
HYADES-B3 & 2002-11-06 & 4:22:31.05 & 17:04:31.0 & 0.7 \\
HYADES-B4 & 2002-11-06 & 4:21:31.11 & 16:31:38.0 & 0.8 \\
HYADES-B5 & 2002-11-06 & 4:21:11.31 & 16:05:48.4 & 0.6 \\
HYADES-B6 & 2002-11-06 & 4:21:40.84 & 15:39:44.6 & 0.6 \\
HYADES-B7 & 2002-11-07 & 4:21:41.61 & 15:21:27.5 & 0.7 \\
HYADES-B8 & 2002-11-07 & 4:21:41.14 & 14:39:40.4 & 0.8 \\
HYADES-B9 & 2002-11-07 & 4:24:25.86 & 14:29:41.7 & 0.7 \\
HYADES-B10 & 2002-11-07 & 4:24:25.96 & 14:04:43.5 & 0.7 \\
HYADES-B11 & 2002-11-07 & 4:27:11.16 & 14:19:46.3 & 0.6 \\
HYADES-B12 & 2002-11-08 & 4:30:00.81 & 14:34:46.3 & 0.7 \\
HYADES-B13 & 2002-11-08 & 4:32:09.73 & 14:49:39.8 & 0.6 \\
HYADES-B14 & 2002-11-08 & 4:32:00.09 & 15:24:44.2 & 0.7 \\
HYADES-B15 & 2002-11-09 & 4:33:31.42 & 15:49:37.3 & 0.7 \\
HYADES-B16 & 2002-11-09 & 4:32:20.49 & 16:19:41.0 & 0.6 \\
HYADES-B17 & 2002-11-10 & 4:31:20.45 & 16:51:34.8 & 0.8 \\
HYADES-B18 & 2002-11-12 & 4:31:00.92 & 17:16:33.2 & 0.6 \\
HYADES-C01 & 2002-11-08 & 4:20:30.91 & 17:29:29.6 & 1.0 \\
HYADES-C02 & 2003-01-26 & 4:23:19.97 & 17:54:38.1 & 0.7 \\
HYADES-C03 & 2003-01-26 & 4:25:20.42 & 18:16:46.3 & 0.7 \\
HYADES-C04 & 2003-01-27 & 4:23:00.30 & 18:41:34.7 & 0.5 \\
HYADES-C05$^\dagger$ & 2003-01-28 & 4:20:29.31 & 17:54:29.9 & 0.8 \\
HYADES-C06 & 2002-12-27 & 4:19:40.61 & 18:21:26.5 & 0.6 \\
HYADES-C07 & 2003-01-28 & 4:19:50.05 & 17:04:46.1 & 0.6 \\
HYADES-C08 & 2003-01-28 & 4:18:10.61 & 16:39:40.4 & 0.7 \\
HYADES-C09 & 2002-09-07 & 4:17:06.06 & 17:02:16.8 & 0.6 \\
HYADES-C10 & 2002-09-07 & 4:17:40.56 & 17:27:27.1 & 0.6 \\
HYADES-C13 & 2002-12-04 & 4:27:15.45 & 13:54:16.5 & 0.9 \\
HYADES-C14 & 2002-12-04 & 4:30:00.29 & 14:09:24.2 & 0.8 \\
HYADES-C15 & 2002-12-04 & 4:32:45.17 & 14:23:26.2 & 0.7 \\
HYADES-C16 & 2002-12-04 & 4:35:45.31 & 14:59:29.8 & 0.8 \\
HYADES-C17 & 2002-12-04 & 4:36:09.91 & 15:34:27.0 & 0.6 \\
HYADES-C18 & 2002-12-04 & 4:26:19.49 & 13:29:23.0 & 0.7 \\
HYADES-C19 & 2002-12-06 & 4:29:00.53 & 13:37:29.3 & 0.6 \\
HYADES-C20 & 2002-12-04 & 4:32:59.91 & 13:57:22.3 & 0.8 \\
HYADES-C21 & 2002-12-05 & 4:35:29.56 & 14:33:15.5 & 0.6 \\
HYADES-C22 & 2002-12-06 & 4:30:29.97 & 13:12:25.9 & 0.7 \\
HYADES-C23 & 2002-12-06 & 4:33:15.01 & 13:32:31.7 & 0.8 \\
HYADES-C24 & 2002-12-06 & 4:35:28.15 & 14:07:29.3 & 0.5 \\
\hline                                   %inserts single line
\multicolumn{5}{l}{$^\dagger$ No short exposure was obtained for this
  field.}\\
\end{tabular}
\end{table}

\section{Finding charts}

We provide here finding charts for the 2 detected T-dwarfs in the
Hyades cluster. The finding charts are a subset of the CFHT12K I-band
images.

   \begin{figure}[h]
     \centering
      \resizebox{0.45\hsize}{!}{\includegraphics[width=4cm]{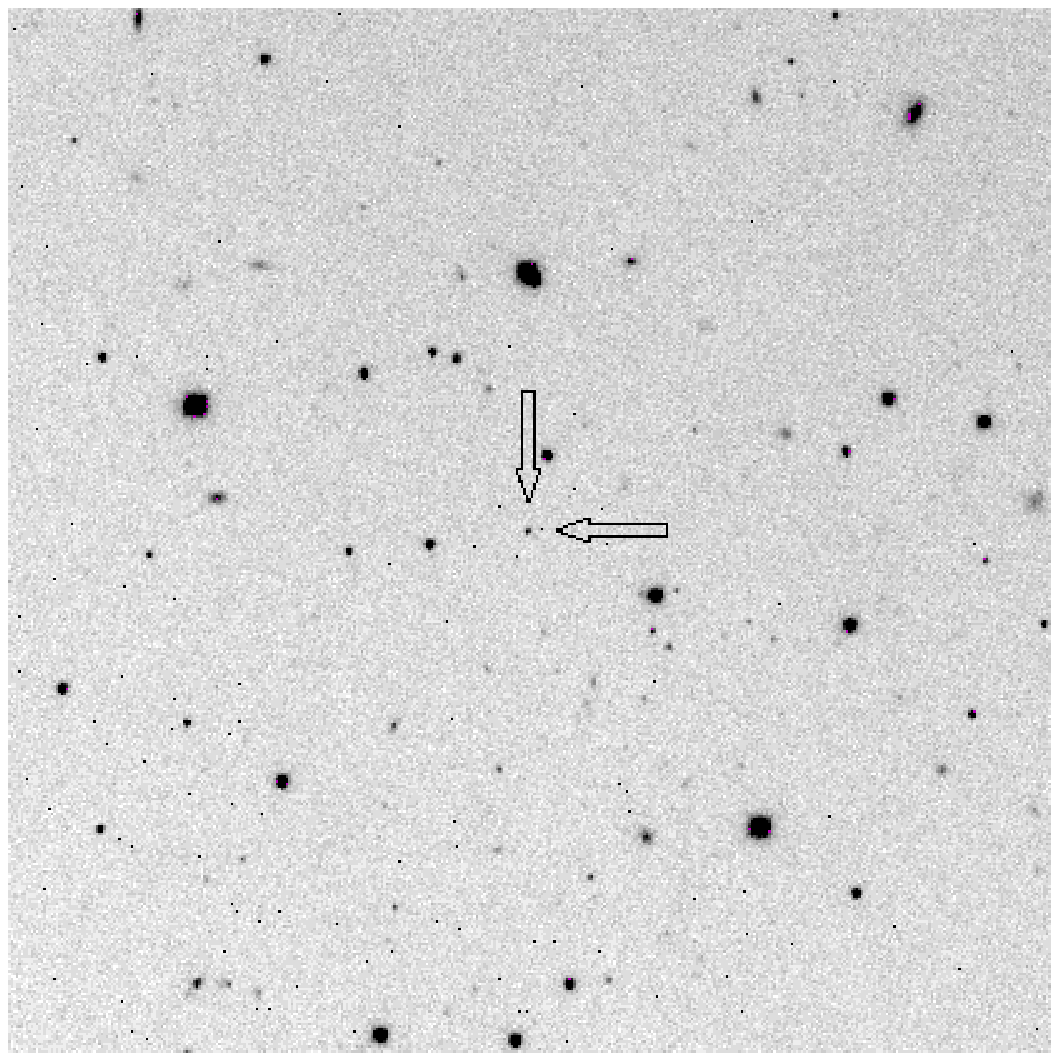}}
      \resizebox{0.45\hsize}{!}{\includegraphics[width=4cm]{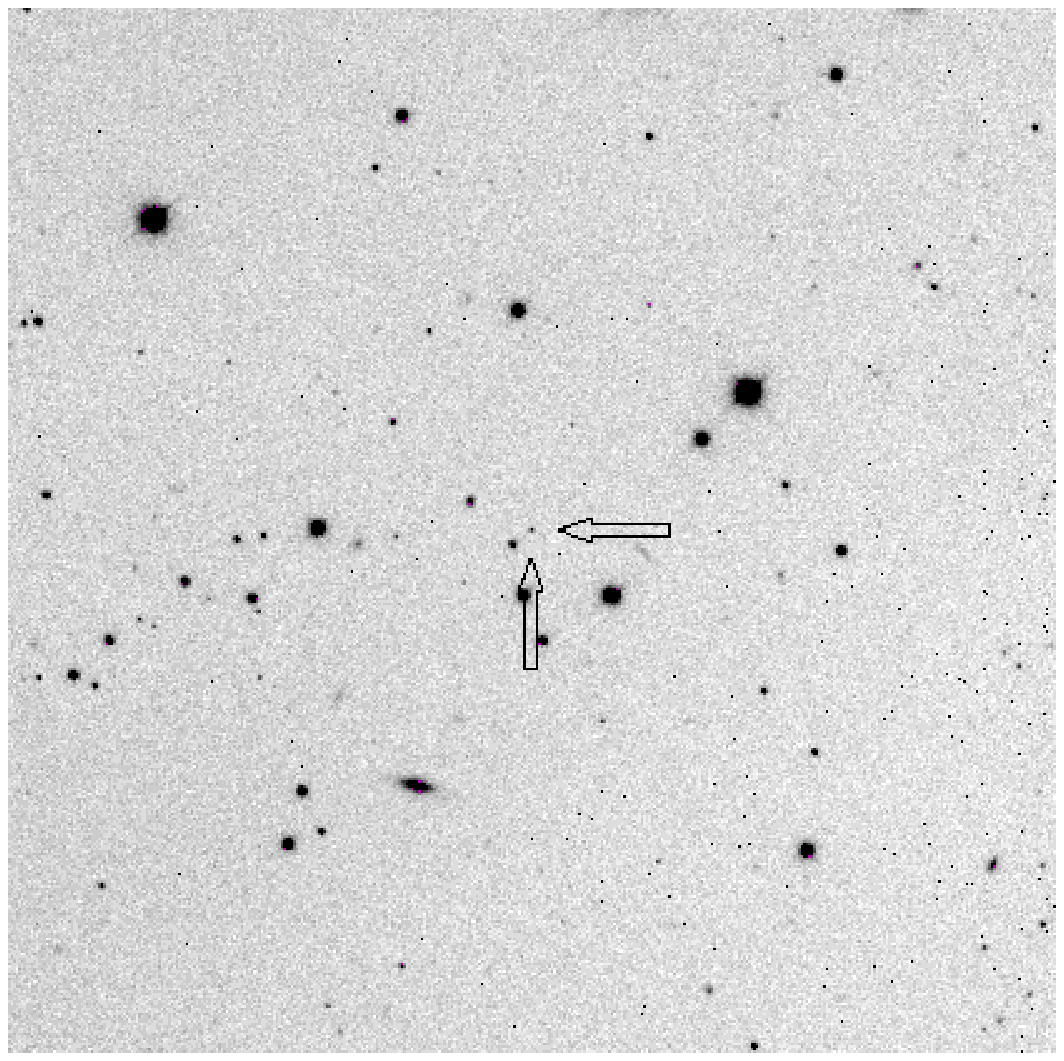}}
      \caption{I-band finding charts for CFHT-Hy-20 (left) and
      CFHT-Hy-21 (right). The size is 3$\times$3$\arcmin$, North is
      up, East is left.}
         \label{fc}
   \end{figure}
%   \begin{figure}[h]
%      \includegraphics[width=7cm]{FC_HA9Ic04_1346l_sub.ps}
%      \caption{I-band finding chart for CFHT-Hy-21. The size is
%      3$\times$3$\arcmin$, North is up, East is left.}
%         \label{App1}
%   \end{figure}

%
%
\section{ Estimating the mass of probable Hyades members in the
Prosser \& Stauffer database. }

The Prosser \& Stauffer database lists 521 probable Hyades members. In
order to combine this database with our dataset and build the complete
mass function for the cluster, we derived the mass of the P\&S
database probable members from their V-band magnitude (with
DM=3.33). For objects with a mass larger than 0.6$M_\odot$, we used
the Mass-M$_V$ relationship derived by Pinsonneault et al. (2004) for
the Hyades. The Mass-M$_V$ relationship spans the mass range from
0.636 to 2.235~M$_\odot$, corresponding to M$_V$ from 1.10 to 8.50
(and V from 4.43 to 11.83). For objects more massive than
2.2~M$_\odot$, we linearily extrapolated Pinsonneault et al.'s (2004)
Mass-M$_V$ relationship, up to the most massive Hyades probable member
listed in the P\&S database with a mass of 2.67~M$_\odot$.

For objects less massive than 0.6~M$_\odot$, we used the
mass-magnitude relationships provided by the Lyon models (Baraffe et
al. 1998; Chabrier et al. 2000). Down to 0.08~M$_\odot$, we used the
NextGen 600~Myr isochrone, while for lower mass objects we used the
Dusty 600~Myr isochrone. The NextGen to Dusty transition is thus taken
to arise at 0.08M$_\odot$, corresponding to an effective temperature
of 2500K. It is usually acknowledged that the Lyon models are not
totally reliable in the V-band, due to missing opacity sources. Hence,
we prefer to derive the mass of the candidates from their I-band
magnitude, since models are believed to be more reliable at this
wavelength. However, while all probable members listed in the P\&S
database have a measured V-band magnitude, only a fraction of them
have a known I-band magnitude. We thus proceeded in 2 steps.

\begin{figure}[h]
   \centering
      \resizebox{0.9\hsize}{!}{\includegraphics{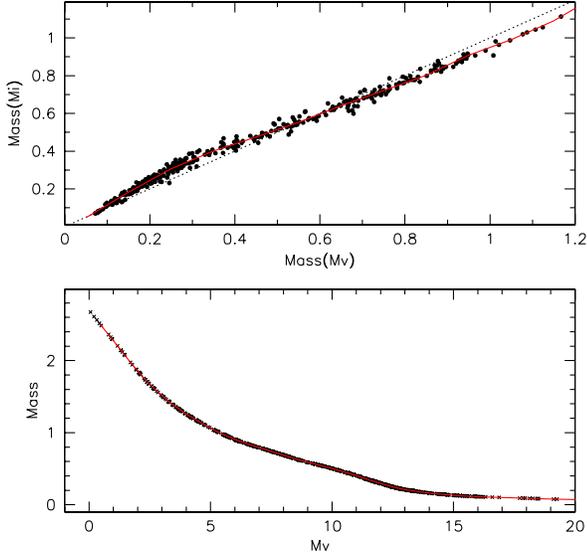}}
      \caption{{\it Top~:} Comparison of the masses estimated from the Lyon
      models using either the Mass-M$_V$ (x-axis) or the Mass-M$_I$
      (y-axis) 600~Myr isochrone. The dotted line shows the locus of
      equal masses while the solid line is a spline interpolation of
      the data. {\it Bottom~:} The Mass-M$_V$ relationship (solid line) used to derive
      mass from the V-band magnitudes of probable Hyades members from
      the Prosser \& Stauffer database (crosses). Above 0.6~M$_\odot$,
      the relationship is that of Pinsonneault et al. (2004), below
      0.6~M$_\odot$ it is derived from the Lyon models (see text).}
         \label{massmv}
   \end{figure}

We first estimated the masses of a subsample of 356 candidates in the
database that have both V- and I-magnitudes. Two estimates of the mass
is obtained for each object, one from the Mass-M$_V$ isochrone, the
other from the Mass-M$_I$ isochrone. Masses derived in both ways are
compared in Figure~\ref{massmv} over the mass range from 0.07 to
1.2~M$_\odot$. It is seen that masses derived from M$_V$ tends to be
systematically lower than those derived from M$_I$ by up to 20\% at
masses less than 0.6~M$_\odot$, while they become slightly larger, by
up to 5\%, for more massive objects. We thus used these mass estimates
to compute the correction to be applied to the V-band mass in order to
convert it to an I-band mass. The correction is shown as the spline
fit interpolation in Fig.~\ref{massmv}. We then simply proceed by
deriving the mass from the V magnitudes of all the low mass P\&S database
probable members and applying the correction computed above to
transform it to an I-band model mass.

The resulting Mass-M$_V$ relationship we use over the whole mass range of
Hyades probable members is shown in Figure~\ref{massmv} and is listed in
Table~\ref{mvmass}. This Mass-Mv relationship combines Pinsonneault et
al.'s (2004) one for masses larger than 0.6~M$_\odot$ and the Lyon
models for lower mass objects.

%----------------------------------------------------------- Fig.5
% Fig.massmv = /home/jbouvier/HyadesABC/RadDist/MF/MassEstimates

%
%______________________________________________________________
%----------------------------------------------------------- Fig.5
% Fig.App2 = /gaglin6/data2/HyadesABC/RadDist/MF/MassEstimates
%
%   \begin{figure}[h]
%   \centering
%   \resizebox{0.9\hsize}{!}{\includegraphics{Mv_Mass_relationship.ps}}
%   \includegraphics{Mv_Mass_relationship.ps}
%   \caption{The Mass-M$_V$ relationship (solid line) used to derive
%      mass from the V-band magnitudes of probable Hyades members from
%      the Stauffer \& Prosser database (crosses). Above 0.6~M$_\odot$,
%      the relationship is that of Pinsonneault et al. (2004), below
%      0.6~M$_\odot$ it is derived from the Lyon models (see text).}      
%         \label{App2}
%   \end{figure}
%
%----------------------------------------------------------- Tab.Appendix

\begin{table}[h]
\caption{The Hyades M$_V$-Mass relationship used in this study.}             % title of Table
\label{mvmass}      % is used to refer this table in the text
\centering                          % used for centering table
\begin{tabular}{l l l l l l l l}        % centered columns (4 columns)
\hline\hline                 % inserts double horizontal lines
M$_V$  &  Mass  & M$_V$  &  Mass  & M$_V$  &  Mass  & M$_V$  &  Mass  \\
& (M$_\odot$) && (M$_\odot$) &&  (M$_\odot$) &&  (M$_\odot$) \\
\hline                        % inserts single horizontal line
23.25 & 0.050 & 14.68 & 0.141 & 11.08 & 0.397 & 4.69 & 1.119 \\
21.66 & 0.056 & 14.11 & 0.158 & 10.61 & 0.446 & 4.01 & 1.256 \\
21.00 & 0.063 & 13.65 & 0.177 & 10.04 & 0.500 & 3.37 & 1.409 \\
20.00 & 0.071 & 13.22 & 0.199 & 9.34 & 0.561 & 2.77 & 1.581 \\
18.99 & 0.079 & 12.81 & 0.223 & 8.56 & 0.630 & 2.21 & 1.774 \\
18.06 & 0.089 & 12.46 & 0.251 & 7.84 & 0.706 & 1.67 & 1.991 \\
17.03 & 0.100 & 12.16 & 0.281 & 7.01 & 0.792 & 1.10 & 2.233 \\
16.06 & 0.111 & 11.82 & 0.316 & 6.12 & 0.889 & 0.45 & 2.506 \\
15.26 & 0.126 & 11.44 & 0.354 & 5.38 & 0.998 &  &  \\
\hline
\end{tabular}
\end{table}

%\end {appendix}

%\begin{appendix}

\section{Hyades probable non-members.  }

We provide here a list of optically-selected Hyades candidates from the (I,
  I-z) CMD which turn
  out to be non members based on follow up K-band photometry and
  proper motion measurement.

%\begin{longtable}{llllll}
%\caption{\label{candidates} Optically selected Hyades candidate members.}\\             % title of Table
%\hline\hline
%Name & RA & Dec. & I & I-z & Notes\\    % table heading 
%&(2000)&(2000)\\
%\hline                        % inserts single horizontal line
%\endfirsthead
%\caption{continued.}\\
%\hline\hline
%Name & RA & Dec. & I & I-z & Notes\\    % table heading 
%&(2000)&(2000)\\
%\hline
%\endhead
%\hline

\begin{longtable}{lllllllll}
\caption{\label{wrong} Hyades probable non-members : candidate members selected from the (I-,
  I-z) CMD but whose infrared colors and/or proper motion is inconsistent
  with membership.}\\             % title of Table
\hline\hline      % is used to refer this table in the text
Name & RA(2000) & Dec(2000) & I & I-z & I-K & $\mu_{\alpha\cos\delta}$& $\mu_\delta$ & Other name\\    % table heading 
&&&&&& \multicolumn{2}{l}{($mas.yr^{-1}$)}  \\
\hline
\endfirsthead
\caption{continued.}\\
\hline\hline
Name & RA(2000) & Dec(2000) & I & I-z & I-K & $\mu_{\alpha\cos\delta}$& $\mu_\delta$ & Other name\\    % table heading 
&&&&&& \multicolumn{2}{l}{($mas.yr^{-1}$)}  \\
\hline      
\endhead
\hline                  % inserts single horizontal line
HA7Ic02-222s  &  4 28 14.2  &  16 58 0  &  12.96  &  0.46  &  2.22  &  26  &  -5  & VA~477\\
HA2Ic07-156s  &  4 25 56.4  &  16 13 8  &  13.58  &  0.52  &  2.35  &  7  &  -21  & \\
HC13c03-570s  &  4 27 18.6  &  13 55 8  &  13.6  &  0.5  &  2.37  &  -46  &  -12  & \\
HB18c03-31s  &  4 31 3.5  &  17 29 39  &  13.76  &  0.62  &  2.52  &  -37  &  -59  & \\
HA3Ic03-197s  &  4 23 57  &  16 32 44  &  13.83  &  0.53  &  2.33  &  1  &  0  & \\
HA2Ic07-287s  &  4 25 43.6  &  16 16 21  &  13.87  &  0.56  &  2.44  &  24  &  -37  & \\
HA1Ic07-119s  &  4 26 25.8  &  15 45 3  &  13.92  &  0.72  &  3.13  &  -25  &  17  & \\
HB1Ic04-70s  &  4 26 53.3  &  17 36 48  &  13.94  &  0.53  &  2.25  &  47  &  -4  & \\
HA5Ic02-407s  &  4 29 59.1  &  16 2 59  &  14.08  &  0.54  &  2.31  &-117  &  -235  & VA~562,~LP~415-143\\
HA9Ic06-243s  &  4 27 4.7  &  15 24 29  &  14.14  &  0.54  &  2.32  &  -43  &  -14  & \\
HB18c02-680s  &  4 30 41.8  &  17 19 43  &  14.18  &  0.74  &  2.83  &  26  &  8  & \\
HC02c09-531s  &  4 23 25.8  &  17 53 44  &  14.25  &  0.55  &  2.51  &  149  &  -139  & \\
HB2Ic01-63s  &  4 26 40.7  &  18 4 17  &  14.27  &  0.6  &  2.51  &  59  &  -132  & \\
HB18c09-393s  &  4 31 12.4  &  17 9 42  &  14.3  &  0.68  &  2.73  &  50  &  -87  & \\
HB11c08-273s  &  4 26 47.9  &  14 11 58  &  14.36  &  0.57  &  2.37  &  69  &  -79  & \\
HC14c01-191s  &  4 29 19.9  &  14 19 36  &  14.53  &  0.67  &  2.67  &  5  &  6  & \\
HC09c11-32s  &  4 18 21.4  &  16 51 34  &  14.58  &  0.6  &  2.61  &  -27  &  -64  & \\
HB3Ic09-301s  &  4 22 39  &  16 57 50  &  14.74  &  0.59  &  2.42  &  9  &  -24  & \\
HB3Ic10-505s  &  4 23 27.7  &  17 2 27  &  14.78  &  0.63  &  2.5  &  9  &  -1  & \\
HA9Ic09-44s  &  4 28 58.6  &  15 17 37  &  14.86  &  0.69  &  2.76  &  77  &  -104  & \\
HB7Ic04-145s  &  4 22 14.1  &  15 30 52  &  15.36  &  0.78  &  4.03  &  25  &  -6  & h4046b\\
HB1Ic08-643s  &  4 25 39.3  &  17 23 2  &  15.56  &  0.72  &  2.66  &  20  &  -5  & \\
HB2Ic01-214s  &  4 26 32.1  &  18 0 27  &  15.56  &  0.75  &  2.9  &  -9  &  8  & \\
HA10c04-89s  &  4 25 14.2  &  15 41 8  &  17.42  &  1.01  &  3.94  &  75  &  -120  & \\
HB1Ic01-386l  &  4 25 23.1  &  17 35 14  &  17.54  &  1.07  &  3.79  &  -2  &  -11  & LH~0422+17,~LH~994\\
HC21c04-2512l  &  4 36 3.1  &  14 35 37  &  18.21  &  1.01  &  3.96  &  21  &  -52  & \\
HB13c09-22s  &  4 32 25.8  &  14 35 44  &  18.77  &  1.06  &  3.41  &  15  &  -49  & \\
HC10c00-186l  &  4 16 21.6  &  17 42 48  &  18.84  &  1.03  &  3.48  &  28  &  -17  & \\
HC09c10-1308l  &  4 17 48.4  &  16 55 59  &  19.05  &  1.13  &  2.8  &  15  &  9  & \\
HC24c02-487s  &  4 35 10.4  &  14 12 51  &  19.13  &  1.11  &  4.04  &  32  &  -13  & \\
HB4Ic04-257s  &  4 22 4.5  &  16 39 44  &  19.18  &  1.03  &  3.95  &  27  &  -15  & \\
HC09c10-2650l  &  4 17 41.1  &  17 1 14  &  19.21  &  1.19  &  3.73  &  22  &  -20  & \\
HB7Ic11-463s  &  4 23 4.7  &  15 18 57  &  19.38  &  1.08  &  4.18  &  40  &  -31  & h7334b\\
HB15c00-4l  &  4 32 9.9  &  16 3 28  &  19.67  &  1.07  &  3.85  &  -7  &  7  & \\
HA2Ic04-474l  &  4 27 22  &  16 34 13  &  19.76  &  1.07  &  3.8  &  -17  &  -2  & \\
HC21c07-1526l  &  4 34 57.4  &  14 28 8  &  19.8  &  1.09  &  5.33  &  -5  &  -23  & \\
HC03c03-516s  &  4 25 27.2  &  18 24 21  &  19.84  &  1.2  &  5.23  &  -14  &  -71  & \\
HC22c00-1508l  &  4 29 19.3  &  13 17 21  &  19.85  &  1.12  &  3.96  &  -31  &  13  & \\
HB6Ic01-25s  &  4 20 49.5  &  15 53 33  &  19.86  &  1.06  &  3.84  &  24  &  -15  & \\
HA9Ic02-41s  &  4 28 30.1  &  15 43 7  &  19.89  &  1.1  &  4.15  &  15  &  -60  & \\
HB8Ic08-171l  &  4 21 17.2  &  14 26 54  &  19.9  &  1.08  &  3.84  &  -9  &  -8  & \\
HA12c07-101l$^\dagger$  &  4 26 12.8  &  14 50 29  &  20  &  1.07  &  3.53  &  90  &  -19  & \\
HB12c04-594l  &  4 30 35.3  &  14 45 32  &  20.11  &  1.08  &  4.34  &  3  &  -49  & \\
HB12c10-1327l  &  4 30 52  &  14 26 41  &  20.19  &  1.1  &  3.73  &  -11  &  2  & \\
HB2Ic07-287l  &  4 26 50.2  &  17 39 30  &  20.28  &  1.11  &  4.41  &  -13  &  -55  & \\
HC15c00-530l  &  4 31 28.7  &  14 35 13  &  20.29  &  1.15  &  3.98  &  -14  &  0  & \\
HA1Ic07-930l  &  4 26 5.2  &  15 49 26  &  20.3  &  1.12  &  4.23  &  46  &  -30  & \\
HC15c09-213l  &  4 33 8.9  &  14 10 40  &  20.45  &  1.11  &  4.07  &  13  &  -26  & \\
HB7Ic04-1273l  &  4 22 19.2  &  15 25 54  &  20.78  &  1.12  &  5.05  &  -11  &  9  & \\
HC21c11-74l  &  4 36 46.1  &  14 19 37  &  20.79  &  1.13  &  3.91  &  21  &  10  & \\
HA9Ic02-266l  &  4 28 27.5  &  15 42 23  &  20.89  &  1.12  &  4.81  &  2  &  -1  & \\
HA10c04-641l  &  4 25 24  &  15 39 41  &  20.99  &  1.12  &  3.43  &  20  &  -50  & \\
HA3Ic01-1181l  &  4 23 8.3  &  16 29 30  &  21.02  &  1.13  &  4.23  &  2  &  13  & \\
HB17c07-1416l  &  4 30 29.7  &  16 44 45  &  21.09  &  1.12  &  4.62  &  32  &  49  & \\
HC10c01-87l  &  4 16 50.2  &  17 42 55  &  21.17  &  1.12  &  4.2  &  0  &  13  & \\
HA1Ic09-460l  &  4 27 16.5  &  15 45 33  &  21.19  &  1.19  &  4.41  &  -26  &  -13  & \\
HC14c01-607l  &  4 29 22.7  &  14 19 10  &  21.36  &  1.2  &  4.69  &  -25  &  -21  & \\
HB7Ic10-1289l  &  4 22 35.3  &  15 15 24  &  21.38  &  1.18  &  4.43  &  -5  &  13  & \\
HB2Ic08-992l  &  4 27 1.8  &  17 44 54  &  21.42  &  1.18  &  4.76  &  -32  &  -37  & \\
HB13c11-245l  &  4 33 25.4  &  14 36 28  &  21.5  &  1.18  &  4.5  &  6  &  -16  & \\
HB3Ic11-684l  &  4 23 46.7  &  16 53 39  &  21.51  &  1.2  &  4.19  &  13  &  -16  & \\
HB17c06-999l  &  4 30 13.2  &  16 43 9  &  21.55  &  1.19  &  5.46  &  17  &  -7  & \\
HC03c04-2872l  &  4 26 6.5  &  18 20 25  &  21.55  &  1.18  &  4.43  &  20  &  -9  & \\
HB12c02-1504l  &  4 29 45.8  &  14 39 10  &  21.63  &  1.17  &  4.14  &  5  &  -30  & \\
HB13c07-1564l  &  4 31 15.5  &  14 43 56  &  21.69  &  1.3  &  5.1  &  -5  &  -2  & \\
HB1Ic03-1536l  &  4 26 20.8  &  17 28 22  &  21.7  &  1.24  &  4.07  &  48  &  52  & \\
HC06c07-1993l  &  4 19 6.1  &  18 21 15  &  21.87  &  1.23  &  5.19  &  10  &  -34  & \\
HA11c07-803l$^1$ & 4 23 42 & 14 58 43 & 21.89 & 1.15 \\
HC10c10-1796l$^1$ & 4 18 27 & 17 24 39 & 21.93 & 1.19 \\
HB9Ic07-234l  &  4 23 27  &  14 17 23  &  21.94  &  1.23  &  5.04  &  79  &  72  & \\
HC06c02-152l$^1$ & 4 19 22 & 18 34 42 & 22.00 & 1.17 \\
HC19c03-2076l$^1$ & 4 29 21 & 13 40 10 & 22.10 & 1.18 \\
HC02c07-578l  &  4 22 28  &  17 43 55  &  22.13  &  1.22  &  5.32  &  36  &  -26  & \\
HB17c08-2177l$^1$ & 4 31 18 & 16 49 34 & 22.16 & 1.17 \\
HA7Ic07-180l$^1$ & 4 27 34 & 16 37 11 & 22.21 & 1.19 \\
HC13c00-520l  &  4 25 52.9  &  14 6 11  &  22.28  &  1.23  &  4.48  &  1  &  46  & \\
HA6Ic01-2048l$^1$ & 4 29 20 & 16 29 46 & 22.33 & 1.19 \\
HA7Ic02-927l  &  4 28 11.9  &  16 57 33  &  22.33  &  1.23  &  4.03  &  -1  &  11  & \\
HC13c09-1062l$^1$ & 4 27 30 & 13 51 7 & 22.36 & 1.22 \\
HB4Ic08-756l  &  4 21 4.7  &  16 22 45  &  22.36  &  1.26  &  4.73  &  5  &  3  & \\
HC15c10-170l$^1$ & 4 33 31 & 14 9 45 & 22.41 & 1.19 \\
HA13c08-1543l  &  4 29 11.8  &  14 59 46  &  22.49  &  1.26  &  3.45  &  -12  &  6  & \\
HB7Ic04-558l  &  4 22 14.2  &  15 31 18  &  22.51  &  1.27  &  5.86  &  -15  &  14  & \\
HB7Ic08-324l$^1$ & 4 21 33 & 15 9 48 & 22.54 & 1.23 \\
HC06c01-591l$^1$ & 4 18 47 & 18 31 18 & 22.55 & 1.21 \\
HB7Ic11-770l  &  4 22 40.2  &  15 12 50  &  22.55  &  1.49  &  5.61  &  16  &  -46  & \\
HA1Ic09-735l  &  4 27 20.4  &  15 48 0  &  22.56  &  1.27  &  4.64  &  23  &  -12  & \\
HC17c02-1196l  &  4 35 46.8  &  15 39 41  &  22.58  &  1.25  &  4.88  &  -10  &  6  & \\
HB9Ic06-530l$^1$ & 4 22 59 & 14 19 48 & 22.61 & 1.21 \\
HB9Ic09-490l$^1$ & 4 24 49 & 14 18 32 & 22.61 & 1.22 \\
HA5Ic01-385l$^1$ & 4 29 34 & 16 7 26 & 22.67 & 1.22 \\
HB12c11-86l  &  4 31 6.1  &  14 20 46  &  22.68  &  1.4  &  4.39  &  -56  &  12  & \\
HB18c03-2563l$^1$ & 4 31 17 & 17 18 30 & 22.7 & 1.25 \\
HC18c10-169l  &  4 27 14.8  &  13 15 38  &  22.77  &  1.26  &  4.23  &  24  &  12  & \\
HA6Ic02-1202l$^1$ & 4 29 56 & 16 31 31 & 22.78 & 1.22 \\
HB1Ic08-1689l  &  4 25 37.7  &  17 20 13  &  22.78  &  1.25  &  4.59  &  2  &  0  & \\
HA8Ic00-3532l$^1$ & 4 23 50 & 16 50 13 & 22.8 & 1.22 \\
HA7Ic11-2242l  &  4 29 43.6  &  16 46 26  &  22.8  &  1.24  &  4.35  &  -11  &  -35  & \\
HA13c08-1866l  &  4 29 6.1  &  15 0 35  &  22.86  &  1.32  &  3.39  &    &    & \\
HC09c02-1520l  &  4 17 0  &  17 6 43  &  22.91  &  1.26  &  4.69  &  -7  &  -2  & \\
HB6Ic09-322l$^\ddagger$  &  4 21 53.4  &  15 32 27  &  22.92  &  1.25  &  7.55  &  16  &  10  & \\
HC10c00-26l  &  4 16 36.5  &  17 43 26  &  22.97  &  1.26  &  5.18  &  -8  &  3  & \\
HC02c11-341l  &  4 24 42.3  &  17 41 52  &  22.99  &  1.31  &  3.71  &  -29  &  -35  & \\
\hline
\multicolumn{9}{l}{$^1$ This candidate has no follow up K-band
  photometry.}\\
\multicolumn{9}{l}{$^\dagger$ This candidate has a proper motion consistent with
  membership but is too blue in I-K to be a cluster member.}\\
\multicolumn{9}{l}{$^\ddagger$ This extremely red object was also
  observed in the J and H bands, yielding (J-H, H-K)=(2.24, 0.91).}\\
\end{longtable}

\end {appendix}

\end{document}